\input harvmac

%%%%%%%%%%%%%%%%%%%%%%% Definitions %%%%%%%%%%%%%%%%

\def\tQ{\tilde Q}
\def\tq{\tilde q}
\def\tm{\tilde m}
\def\tf{\tilde f}
\def\ov{\overline}
\def\oN{\bar{N}}
\def\ot{\bar{t}}
\def\of{\bar{f}}
\def\oa{\bar{a}}
\def\oV{\bar{V}}
\def\om{\bar{m}}
\def\tilde{\widetilde}
\def\hat{\widehat}
\def\cN{{\cal N}}
\def\L{\Lambda}
\def\diag{{\rm diag}}
\def\cof{{\rm cof}}
\def\const{{\rm const}}

%%%%%%%%%%%%%%%%%%%%%%%%%%%%%%%%%%%%%%%%%%%%%%%%%%%
%%%%%%%%%%%%%%%%% References %%%%%%%%%%%%%%%%%%%%%%

%\SeibergPQ
\lref\SeibergPQ{ N.~Seiberg,
``Electric - magnetic duality in supersymmetric nonAbelian gauge theories,''
Nucl.\ Phys.\ B {\bf
435}, 129 (1995) [arXiv:hep-th/9411149].
%%CITATION = HEP-TH 9411149;%%
}

%\IntriligatorAU
\lref\IntriligatorAU{ K.~A.~Intriligator and N.~Seiberg,
``Lectures on supersymmetric gauge theories and electric-magnetic duality,''
Nucl.\ Phys.\ Proc.\ Suppl.\  {\bf 45BC}, 1 (1996)
[arXiv:hep-th/9509066].
%%CITATION = HEP-TH 9509066;%%
}

%\KonishiHF
\lref\Konishi{ K.~Konishi,
``Anomalous Supersymmetry Transformation Of Some Composite Operators In Sqcd,''
Phys.\ Lett.\ B {\bf 135}, 439 (1984).
%%CITATION = PHLTA,B135,439;%%
%\KonishiTU
K.~i.~Konishi and K.~i.~Shizuya,
``Functional Integral Approach To Chiral Anomalies In Supersymmetric Gauge Theories,''
Nuovo Cim.\ A {\bf 90}, 111 (1985).
%%CITATION = NUCIA,A90,111;%%
}

\lref\Kuta{
D.~Kutasov,
``A Comment on duality in N=1 supersymmetric nonAbelian gauge theories,''
Phys.\ Lett.\ B {\bf 351}, 230 (1995)
[arXiv:hep-th/9503086].
%%CITATION = HEP-TH 9503086;%%
}

\lref\KutaS{
D.~Kutasov and A.~Schwimmer,
``On duality in supersymmetric Yang-Mills theory,''
Phys.\ Lett.\ B {\bf 354}, 315 (1995)
[arXiv:hep-th/9505004].
%%CITATION = HEP-TH 9505004;%%
}

\lref\KSS{D.~Kutasov, A.~Schwimmer and N.~Seiberg,
``Chiral Rings, Singularity Theory and Electric-Magnetic Duality,''
Nucl.\ Phys.\ B {\bf 459}, 455 (1996)
[arXiv:hep-th/9510222].
%%CITATION = HEP-TH 9510222;%%
}

%\CachazoKX
\lref\cachazo{
F.~Cachazo,
``Notes on supersymmetric Sp(N) theories with an antisymmetric tensor,''
arXiv:hep-th/0307063.
%%CITATION = HEP-TH 0307063;%%
M.~Matone,
``The affine connection of supersymmetric SO(N)/Sp(N) theories,''
JHEP {\bf 0310}, 068 (2003)
[arXiv:hep-th/0307285].
%%CITATION = HEP-TH 0307285;%%
}

%\DijkgraafFC
\lref\DV{R.~Dijkgraaf and C.~Vafa,
``Matrix models, topological strings, and supersymmetric gauge theories,''
Nucl.\ Phys.\ B {\bf 644}, 3 (2002)
[arXiv:hep-th/0206255];
%%CITATION = HEP-TH 0206255;%%
%\DijkgraafVW
``On geometry and matrix models,''
Nucl.\ Phys.\ B {\bf 644}, 21 (2002)
[arXiv:hep-th/0207106];
%%CITATION = HEP-TH 0207106;%%
%\DijkgraafDH
 ``A perturbative window into non-perturbative physics,''
arXiv:hep-th/0208048.
%%CITATION = HEP-TH 0208048;%%
}

%\CachazoRY
\lref\CDSW{ F.~Cachazo, M.~R.~Douglas, N.~Seiberg and E.~Witten,
``Chiral rings and anomalies in supersymmetric gauge theory,''
JHEP {\bf 0212}, 071 (2002)
arXiv:hep-th/0211170.
%%CITATION = HEP-TH 0211170;%%
}

\lref\nati{
%\SeibergJQ
N.~Seiberg,
``Adding fundamental matter to 'Chiral rings and anomalies in  supersymmetric gauge theory',''
JHEP {\bf 0301}, 061 (2003)
[arXiv:hep-th/0212225].
%%CITATION = HEP-TH 0212225;%%
}
\lref\CSW{
F.~Cachazo, N.~Seiberg and E.~Witten,
``Chiral Rings and Phases of Supersymmetric Gauge Theories,''
JHEP {\bf 0304}, 018 (2003)
[arXiv:hep-th/0303207].
%%CITATION = HEP-TH 0303207;%%
}

%\AffleckMK
\lref\ADS{
I.~Affleck, M.~Dine and N.~Seiberg,
``Dynamical Supersymmetry Breaking In Supersymmetric QCD,''
Nucl.\ Phys.\ B {\bf 241}, 493 (1984);
%%CITATION = NUPHA,B241,493;%%
%\AffleckXZ
``Dynamical Supersymmetry Breaking In Four-Dimensions And Its Phenomenological Implications,''
Nucl.\ Phys.\ B {\bf 256}, 557 (1985).
%%CITATION = NUPHA,B256,557;%%
}

\lref\IW{
K.~Intriligator and B.~Wecht,
``The exact superconformal R-symmetry maximizes $a$,''
Nucl.\ Phys.\ B {\bf 667}, 183 (2003)
arXiv:hep-th/0304128.
%%CITATION = HEP-TH 0304128;%%
}

\lref\KPS{
D.~Kutasov, A.~Parnachev and D.~A.~Sahakyan,
``Central charges and $U(1)_R$ symmetries in ${\cal N} = 1$ super Yang-Mills,''
JHEP {\bf 0311}, 013 (2003)
arXiv:hep-th/0308071.
%%CITATION = HEP-TH 0308071;%%
}

%\ArgurioYM
\lref\Ferretti{
R.~Argurio, G.~Ferretti and R.~Heise,
``An introduction to supersymmetric gauge theories and matrix models,''
Int.\ J.\ Mod.\ Phys.\ A {\bf 19}, 2015 (2004)
[arXiv:hep-th/0311066].
%%CITATION = HEP-TH 0311066;%%
}

\lref\feng{
B. Feng,
``Seiberg duality in matrix model,''
[arXiv:hep-th/0211202].
%%CITATION = HEP-TH 0211202;%%
B.~Feng and Y.~H.~He,
``Seiberg duality in matrix models. II,''
Phys.\ Lett.\ B {\bf 562}, 339 (2003)
[arXiv:hep-th/0211234].
%%CITATION = HEP-TH 0211234;%%
B.~Feng,
``Note on Seiberg duality in matrix model,''
Phys.\ Lett.\ B {\bf 572}, 68 (2003)
[arXiv:hep-th/0303144].
%%CITATION = HEP-TH 0303144;%%
}

%\CsakiVV
\lref\Csaki{
C.~Csaki and H.~Murayama,
``Instantons in partially broken gauge groups,''
Nucl.\ Phys.\ B {\bf 532}, 498 (1998)
[arXiv:hep-th/9804061];
%%CITATION = HEP-TH 9804061;%%
%\CsakiFM
``New confining N = 1 supersymmetric gauge theories,''
Phys.\ Rev.\ D {\bf 59}, 065001 (1999)
[arXiv:hep-th/9810014].
%%CITATION = HEP-TH 9810014;%%
}

\lref\Cava{
F.~Cachazo and C.~Vafa,
``N = 1 and N = 2 geometry from fluxes,''
arXiv:hep-th/0206017.
%%CITATION = HEP-TH 0206017;%%
}

%\KleinWE
\lref\Klein{
M.~Klein and S.~J.~Sin,
``Matrix model, Kutasov duality and factorization of Seiberg-Witten curves,''
arXiv:hep-th/0310078.
%%CITATION = HEP-TH 0310078;%%
}

%\AhnYM
\lref\masaki{
C.~h.~Ahn, B.~Feng, Y.~Ookouchi and M.~Shigemori,
``Supersymmetric gauge theories with flavors and matrix models,''
arXiv:hep-th/0405101.
%%CITATION = HEP-TH 0405101;%%
}

%\AldayGB
\lref\aldacira{
L.~F.~Alday and M.~Cirafici,
``Effective superpotentials via Konishi anomaly,''
JHEP {\bf 0305}, 041 (2003)
[arXiv:hep-th/0304119].
%%CITATION = HEP-TH 0304119;%%
}

\lref\JB{
J.~H.~Brodie,
``Duality in supersymmetric $SU(N_c)$ gauge theory with two adjoint chiral superfields,''
Nucl.\ Phys.\ B {\bf 478}, 123 (1996)
[arXiv:hep-th/9605232].
%%CITATION = HEP-TH 9605232;%%
}

%%%%%%%%%%%%%%%%%%%%%%%%%%%%%%%%%%%%%%%%%%%%%%%%%%%%

\Title{\vbox{ \hbox{SISSA/62/2004/EP}  }}
{\vbox{\centerline{Chiral Rings, Anomalies}
\medskip
\centerline{and Electric--Magnetic Duality} }}
\smallskip
\centerline{Luca Mazzucato}
\smallskip
\bigskip
\centerline{International School for Advanced Studies
(SISSA/ISAS)} \centerline{\it Via Beirut 2 - 4, 34014 Trieste,
Italy, and INFN, sez. di Trieste, Italy}
\medskip

\bigskip
\vskip 1cm

\noindent We study electric--magnetic duality in the chiral ring of
a supersymmetric $U(N_c)$ gauge theory with adjoint and fundamental matter, in presence of
a general confining phase superpotential for the adjoint and the mesons. We find the
magnetic solution corresponding to both the pseudoconfining and higgs electric vacua.
By means of the Dijkgraaf--Vafa method, we match the effective
glueball superpotentials and show that in some
cases duality works exactly offshell. We give also
a picture of the analytic structure of the resolvents
in the magnetic theory, as we smoothly interpolate between different
higgs vacua on the electric side.

\vskip 0.5cm

\Date{August 2004}

\newsec{Introduction and Summary}

Supersymmetric $\cN=1$ gauge theories with matter exhibit a generalization of electric--magnetic duality to the case of non--abelian gauge groups, as found by Seiberg in the nineties \SeibergPQ.
This kind of duality
is  not exact at all scales, but it holds at large distances only
and it proves to be a powerful tool in understanding the infrared dynamics of gauge theories.
Consider an asymptotically free supersymmetric
gauge theory, that we will call {\it electric},
whose renormalization group flow has a fixed
point at a long distance scale, where the physics
is described by a superconformal field theory. The {\it magnetic} dual is another
theory which flows to the same fixed point.
In other words, the physics at the infrared point
is described equivalently by both theories. The two sets of degrees of freedom
of the dual pair might be in general very different at the level of the microscopic
lagrangian, as duality holds only for the two low energy effective field theories.

There are basically two different methods to study
the low energy superpotentials of $\cN=1$ supersymmetric gauge theories.
We can use of the tools of holomorphy
and symmetry to contrain the superpotential part of the low energy effective action,
in such a way to uniquely fix it (for a review see \IntriligatorAU). Alternatively,
in theories with confinement and a mass gap, another way to compute the
effective glueball superpotential
has been proposed by Dijkgraaf and Vafa (DV) \DV\ and further developed by Cachazo et al. \CDSW,
based on the solution of some generalized Konishi anomaly equations.

Let us see how these methods work
in the prototypical example of electric--magnetic duality, namely
Seiberg duality in SQCD with $U(N_c)$
gauge group and $N_f$ flavors of quarks $Q^f$ and antiquarks $\tQ_{\tf}$ \SeibergPQ.
If we want to apply DV, the theory has to be massive and we need a tree
level mass term
\eqn\sqcdel{
W_{tree}=m\tQ_f Q^f.
}
Even if classically the mesons vanish, at the quantum level their expectation value
is set by the Konishi anomaly to $\langle \tQ_{f} Q^f\rangle=N_f S/m$, where $S$
is the glueball superfield \Konishi. The effective glueball superpotential is recovered
by integrating this exact expectation value with respect to the corresponding coupling.
We have to add also possible coupling independent terms, that in this case are the
Veneziano--Yankielowicz superpotential $N_cS(1-\log S)$ and the one--loop
exact renormalization of the gauge field $(3N_c-N_f)S\log\L$, obtaining the
glueball effective superpotential
\eqn\sqcdeff{
W_{eff}=S\left(\log{m^{N_f}\L^{3N_c-N_f}\over S^{N_c}}+N_c\right).
}
The magnetic dual of this theory is a supersymmetric gauge theory
with $U(\oN_c)$ gauge group, $N_f$ flavors of magnetic quarks $q_f$ and
antiquarks $\tq^{\tf}$ and $N_f^2$ gauge singlets $P^f_{\tf}$, that represent
the electric mesons. The classical magnetic superpotential corresponding
to \sqcdel\ is $\ov{W}_{tree}={1\over \mu}P\tq q+\ov{m}\tr P$.
In the magnetic theory we have two basic equations to solve. The first is
the singlet equation of motion, which completely fix the
magnetic mesons to $\langle \tq q\rangle=-\mu \ov{m}$.
Since the singlets are not coupled to the gauge fields, their equations of motion
are exact in the chiral ring of the quantum theory. Then we have the
Konishi anomaly, that sets $\langle P\rangle=-\bar{S}/\ov{m}$.
The effective glueball superpotential is then computed as in the electric case and
we get
\eqn\sqcdmaf{
\ov{W}_{eff}=\bar{S}\left(\log{\bar{S}^{N_f-\oN_c}\tilde\L^{3\oN_c-N_f}\over(-\ov{m}\mu)^{N_f}}+(\oN_c-N_f)\right).
}
To find the duality map, we first match the electric mesons with the magnetic singlets,
since they are directly related by a Legendre transform, and we see that $S=-\bar{S}$
and $m=\ov{m}$. Then we match the effective glueball superpotential
and find the relation between the gauge groups $\oN_c=N_f-N_c$ as well as the
scale matching relation $\L^{3N_c-N_f}\tilde\L^{3\oN_c-N_f}=(-)^{N_f-N_c}\mu^{N_f}$.

However, the DV method is not really necessary in this case.\foot{The matrix model
approach to Seiberg duality has been first used in \feng.} We can easily obtain
the onshell expectation values of chiral operators by studying
the nonperturbative low energy superpotentials of electric and magnetic theories,
without ever introducing the glueball superfield.
On the other hand, we can also integrate in the glueball superfield to obtain directly
the glueball effective
superpotential. On the electric side, the low energy theory
is just pure $U(N_c)$ SYM, whose nonperturbative superpotential is
$W_{low}=N_c(\L_{low}^{3N_c})^{1\over N_c}$,
which is responsible for gaugino condensation.
One first matches the low energy scale $\L_{low}^{3N_c}=m^{N_f}\L^{3N_c-N_f}$ and then
just integrate in the glueball to obtain directly \sqcdeff. On the other side, in the
magnetic theory the singlet equations of motion force all the flavors to be higgsed,
thus the low energy theory is pure SYM with gauge group $U(\oN_c-N_f)$, whose low energy
superpotential is the same as the electric one but with the appropriate magnetic quantities instead.
By matching the magnetic scales
%$\tilde\L_{low}^{3(\oN_c-N_f)}=\tilde\L^{3\oN_c-N_f}/\langle\tq q\rangle^{N_f}$
and again integrating in the glueball we obtain \sqcdmaf.

\subsec{Duality in SQCD with Adjoint Matter}

In this paper we will study a more complicated version of Seiberg duality,
first analyzed by Kutasov, Schwimmer and Seiberg
(KSS), namely
a supersymmetric gauge theory with gauge group $U(N_c)$
and $N_f$ flavors of quarks
$Q^f$ and antiquarks $\tQ_{\tf}$ and a chiral superfield $X$ in the adjoint
representation of the gauge group \Kuta\KutaS\KSS. The magnetic dual of the theory
without a superpotential is not known. But we can study deformations
by relevant superpotential couplings, for which we know the dual theory.
A way to simplify
the dynamics, which was studied by KSS, is to add a generic polynomial
superpotential for the adjoint\foot{These operators
are usually referred to as dangerously irrelevant, meaning that
they are irrelevant at the UV fixed point when $n\geq3$, but they become
relevant as we flow to the infrared.}
\eqn\confefi{\eqalign{
W_{el}=&\Tr V(X),\cr
V(z)=&\sum_{k=1}^n{t_k\over k+1}x^{k+1},
}}
that drives the theory to a confining phase in the infrared, leaving
at low energy no dynamics but rather just a discrete set of vacua.

The magnetic dual of the theory \confefi\ is a supersymmetric gauge theory with gauge
group $U(\oN_c)$, where $\oN_c=nN_f-N_c$, and $N_f$ flavors of dual
quarks $q_{\tf}$ and antiquarks $\tq^f$, an adjoint chiral
superfield  $Y$ and $N_f^2$ gauge singlets $(P_j)^f_{\tf}$, $j=1,\ldots,n$, that
represent the electric mesons
$$
P_j=\tQ X^{j-1}Q.
$$
The magnetic theory is defined by the tree level superpotential
\eqn\confema{
W_{mag}=-\Tr V(Y)+\tq\tm(P,Y)q,
}
where $\tm(P,z)$ is a certain degree $n-1$ polynomial, whose coefficients
depend on the gauge singlets $P_j$. This magnetic polynomial will
be the crucial quantity to evaluate in the quantum theory.
Even if classically the chiral rings and the vacua of the two theories are very different,
quantum mechanically they coincide. In particular, KSS proposed
that the dynamically generated scales $\Lambda$ of the electric theory
and $\tilde{\Lambda}$ of the magnetic theory obey the matching
relation
\eqn\scalerel{
\Lambda^{2N_c-N_f}\tilde{\Lambda}^{2\oN_c-
N_f}=\mu^{2N_f}t_n^{-2N_f},
}
which is very similar to the
corresponding scale matching of SQCD we discussed above, and they checked
the matching against various flows.

\subsec{Generic Deformation of KSS and Duality}

The purpose of this paper is to generalize the analysis of KSS by
considering the most
generic electric superpotential, obtained by adding to \confefi\ a meson deformation
\eqn\intrel{
\Tr V(X)+\tilde Q_{\tilde f}m(X)_f^{\tilde f} Q^f,
}
where in the classical chiral ring the degree of the meson polynomial
$m(z)$ is at most $n-1$. At the classical level,
this electric theory presents two different kinds of vacua.
In the first vacuum, that we denote as {\it pseudoconfining}, the fundamentals vanish
and the adjoint acquires a vacuum expectation that drives the theory to
a product of low energy $U(N_i)$ SQCD blocks such that $\sum_{i=1}^n N_i=N_c$.
The other vacuum is called the {\it higgs} vacuum and is characterized
by a nonvanishing classical vev for the fundamentals.\foot{In
presence of matter in the fundamental representation of the gauge group
there is no phase transition between higgs and confining
regimes and in the quantum theory one can continuously interpolate between them.}
In this case the rank of the gauge group decreases. If we higgs $L$ colors,
then the low energy theory is still a product of $U(N_i)$ SQCD blocks, but now
$\sum_{i=1}^n N_i=N_c-L$.

Our first analysis of the duality will focus on the map between the electric
and magnetic classical vacua in both the pseudoconfining and the higgs phase.
The magnetic dual of the theory \intrel\
contains, in addition to the superpotential \confema, a source term for the gauge singlets
$\sum_{k=1}^{\deg m+1}m_kP_k$,
where $m_k$ are the coefficients of $m(z)$. The magnetic vacua will depend then on the details
of the electric meson polynomial: each flavor appearing
in $m(z)^{\tf}_f$
turns on a higgsed block in the magnetic adjoint $\langle Y\rangle$.
In particular, we will study the magnetic vacuum corresponding
to the electric higgs phase, characterized by
a nonzero classical vev for the magnetic singlets $P_j$.
In our classical
solution, as we increase the higgsed
directions in the electric theory, thus driving it to weaker coupling,
the higgsed block in the magnetic theory decreases its rank, driving the dual theory
to stronger coupling.

We will consider then the map between the chiral rings of the two quantum theories.
Due to the presence of a large number of couplings in the tree level action \intrel,
the study of the effective superpotentials by the conventional methods
of holomorphy and symmetries is more involved in this case. Therefore,
we found more convenient to analyze the quantum theory with the DV method, along the lines
discussed above for SQCD. In particular, we will concentrate on
the operators that generate the chiral ring
$$
M(z)=\langle \tQ{1\over z-X}Q\rangle,\qquad T(z)=\langle\Tr{1\over z-X}\rangle.
$$
A generalized version of the Konishi anomaly allows us to solve explicitly
for these operators as functions of the glueball superfield $S$ and the couplings \nati\CSW\
and we can integrate them to obtain the glueball effective superpotentials.
By matching first the electric mesons with the magnetic singlets
and then the two effective superpotentials, we will derive
the relation between the two gauge groups $\oN_c=nN_f-N_c$
as well as the scale matching \scalerel\ and the map between the electric
and magnetic chiral ring operators. The low energy electric and magnetic
theories will be both described by the same
hyperelliptic Riemann surface $y^2=V'(z)^2+f(z)$, a double--sheeted cover of the plane, where the
quantum deformation
$f(z)$ is a degree $n-1$ polynomial.
The pseudoconfining and higgs duality map will turn out
to be rather different, though. In particular, in the electric pseudoconfining phase
the magnetic anomaly equations
are solved by the simple condition
\eqn\fico{
m(a_i)\tm(a_i)=f(a_i),
}
for $i=1,\ldots,n$, where $a_i$ are the roots of $V'(z)$. This condition
will ensure also the match of the electric and magnetic chiral rings and will reproduce
the Konishi anomaly in each low energy SQCD block.

The DV method allows us to study also the rich analytic structure of the
low energy effective theory.
Even if the electric and magnetic theory have
the same curve, the meromorphic functions $M(z)$ and $T(z)$
living on the curve have very different analytic structures on
the two sides. We will picture their analytic behavior as follows.
According to \CSW, an higgs eigenvalue in the electric theory is seen
as a pole of $M(z)$ on the first (semiclassical) sheet of the curve. As we will see,
in the magnetic theory the corresponding $\tilde{M}(z)$ will have
$n-1$ poles on the first sheet. We can
higgs twice the electric theory by bringing
a second pole of $M(z)$ from the second (invisible) sheet into the
first one. The magnetic theory behaves in two different
ways depending on whether we higgs different electric flavors or several
times the same flavor.\foot{At most we can higgs $n-1$ color directions on the same flavor,
corresponding to the degree of the meson polynomial $m(z)$.}
We will see that, in this latter case, the second electric higgsing
corresponds in the magnetic theory to moving one of the $n-1$
poles away from the first into the second sheet.

\subsec{Offshell vs. Onshell Duality}

Seiberg duality describes of an electric and a magnetic theory that, inside the conformal window,
flow to the same fixed point in the infrared, where
both theories are in a
non--abelian Coulomb phase. The dual effective actions for the massless fields
at the fixed point are equivalent. However, a necessary requirement for the
DV method to apply is that all the fields have to be massive:
both theories flow to just a discrete set of vacua in the far infrared. The effective glueball
superpotentials we are computing with the DV method are offshell
actions valid above the energy scale set by the mass of the glueballs.
If the theory has a mass gap, a natural expectation would be that duality works onshell only,
after integrating out all massive fields,
unless we could make sense of an $S$--matrix for the glueballs.
In other words, we would not really expect a change
of variable in the matrix integral, that computes the electric glueball superpotential,
to give us back the magnetic glueball superpotential.
However, at the level of
SQCD we have seen that Seiberg duality works exactly offshell, as will
occur in the first case we will analyze, when the meson
polynomial is just a mass term for all the flavors.
When the meson polynomial is $z$--dependent, instead,
offshell duality will hold only for the first term in the
semiclassical expansion of the theory.

\subsec{Outline of the Paper}

Our main concern will be to compare
electric and magnetic results at every stage of the computation.
For this reason, we will tackle separately the two electric pseudoconfining and
higgs vacua and in each of them we will match
first the classical and then the quantum theories.

In section 2 we will consider the electric pseudoconfining vacuum
in presence of the generic deformation \intrel. We will find
the corresponding classical magnetic solution and see that this is
only valid for a small number of massive electric flavors, due to the presence
of instanton effects in the broken magnetic gauge group. We will then
study the quantum chiral rings of the dual pair by the DV method.
First we will consider the case in which the electric meson superpotential
is just a mass term, and we will show that in this case duality
works exactly offshell. For a generic meson polynomial, instead, the solution
\fico\ that we found is not exact offshell, but still it reproduces
the usual Konishi anomaly in the low energy SQCD blocks and
we believe it to hold onshell.

We will consider then the electric higgs phase in section 3 and
follow the same steps of the previous section, first the classical
and then the quantum analysis, gaining in this way
a complete picture of duality in the different vacua.
Even if the solution of the quantum theory
in this case will be implicit, we will be able to sketch the analytic
behavior of the magnetic resolvents when moving the poles between
the two sheets in the electric theory.

In section 4 we will consider the case of cubic
tree level superpotential to see how duality works in a specific example and
finally, in section 5, we will speculate about some questions raised by our
analysis.

In the appendices we postponed some details of our computation of the
effective superpotential of section 2, which is different from
the one in \CSW. In the last appendix we show a
classical magnetic solution that generalizes the ones in section 3 for the higgs phase.

\newsec{The Electric Pseudoconfining Phase}

In this section we will see how electric--magnetic duality works
in the electric pseudoconfining phase. Our notations will be as follows. In
the classical analysis of sections $2.1$ and $2.2$,
we will always use the electric couplings
to describe the magnetic theory, assuming that we know the duality map.
In the analysis of the quantum theory, from section $2.3$ on,
we will overline the magnetic couplings to avoid possible confusion and
then derive the duality map.

\subsec{The Classical Vacua}

\vskip0.2cm
\centerline{\it The Electric Theory}
\vskip0.2cm

Let us set the stage for our calculations. We will consider an $\cN=1$ supersymmetric gauge theory
with $U(N_c)$ gauge
group, that we will call {\it electric}. The matter content consists of
$N_f$ flavors of quarks $Q^f$ and antiquarks $\tQ_{\tf}$ and a chiral superfield
$X$ in the adjoint representation of the gauge group. We will at first let the
theory flow to its infrared superconformal fixed point. Then we will turn on
the generic
tree level superpotential
\eqn\elth{\eqalign{
W_{el}=&\Tr V(X)+\tilde Q_{\tilde f}m(X)_f^{\tilde f} Q^f,\cr
V'(z)=&\sum_{i=1}^n t_i z^i,\cr
m(z)_f^{\tilde f}=&\sum_{k=1}^{l+1}(m_k)^{\tf}_fz^{k-1},
}}
which is irrelevant in the UV but becomes relevant in the infrared.
It will
be useful to parameterize the adjoint polynomial
as $V'(z)=t_n\prod_{i=1}^n(z-a_i)$ in terms of its roots.
We denote the roots of the meson polynomial $m(z)$ as $x_k$, for $k=1,\ldots,l$.
The degree
of $m(z)$ is $l\leq n-1$, since higher mesons are trivial in the classical chiral
ring, that contains the following
operators
\eqn\meselno{
\Tr X^{j},\qquad \tQ
X^{j-1}Q,
}
for $ j=1,\ldots,n$, as well as operators of the kind $\Tr W_{\alpha}X^j$ and $\Tr W_\alpha W^\alpha X^j$.
However, $W_\alpha Q^f$ and $\tilde Q_{\tilde f} W_\alpha$ are not
in the chiral ring. Also, since the gauge group is $U(N)$ rather than
$SU(N)$ we do not include ``baryonic operators".
Our main attention will be focused on the following chiral operators, that
generate the chiral ring
\eqn\chiralopf{\eqalign{
R(z)&= -{1 \over 32 \pi^2} \Tr {W_\alpha W^\alpha \over
z-X},\cr
M_{\tilde f}^f(z) &= \tilde Q_{\tilde f} {1\over z-X} Q^f,\cr
T(z)&= \Tr {1\over z-X}, \cr
w_\alpha(z)&= {1 \over 4\pi} \Tr {W_\alpha\over z-X}.
}}
We will set to zero
in the following $w_\alpha(z)$ since its duality properties are automatic
and does not constrain the other results.

This theory exhibits two kinds of classically distinct vacua, that we will call
pseudoconfining and higgs vacua. In this section we will be concerned only with the former
and leave
the analysis of the higgs vacuum to the section 3.
The pseudoconfining vacua are characterized by vanishing expectation values for the
fundamentals
\eqn\convac{\eqalign{
X=&\pmatrix{ a_1
&&\cr &        .&\cr &     &       a_n}\cr \tilde Q_{\tilde f}=&0,\quad Q^{f}=0,
}}
where each $a_i$ has multiplicity $N_i$ such that $\sum_iN_i=N_c$. The reason why these are called
``pseudoconfining" rather than ``confining" vacua is that, due to the presence of
fields in the fundamental representation of the gauge group, there is no phase transition
between these vacua and the higgs ones and in the quantum theory they are continuously connected.
At low energy the
theory consists of a set of decoupled $U(N_i)$ SQCD with $N_f$ flavors, while
the adjoint has been integrated out.\foot{If we allow for double roots in $V'(z)$ we end up with
adjoint SQCD with a cubic tree level superpotential for the low energy adjoint superfield. For
simplicity we will consider superpotential with only single roots, though.}
The rank of the gauge group does not decrease along this flow.

%%%%%%%%%%%%%%%%%%%%%%%%%%%%%%%%%%%%%%%%%%%%%%%%%%%%%%%%%%%%%%%%%%%%

\vskip0.2cm
\centerline{\it The Magnetic Theory}
\vskip0.2cm

The magnetic theory corresponding to \elth\ is
again an $\cN=1$ supersymmetric gauge theory with gauge group $U(\oN_c)$
and $N_f$ flavors of dual quarks $q_f$ and antiquarks $\tq^{\tf}$.\foot{Note that $\tq^{\tf}$ is in the
fundamental representation of the flavor symmetry group, while
$q_f$ is in the antifundamental.} We also add a chiral superfield
$Y$ in the adjoint and $N_f^2$ gauge singlets $(P_j)^f_{\tf}$, for $j=1,\ldots,n$.
We first let this theory flow to its
interacting superconformal fixed point, then we add the following superpotential
\eqn\magtre{\eqalign{
W_{mag}=&-\Tr V(Y)
+\tilde q\tilde{m}(P,Y)\, q+\sum_{j=1}^{l+1}m_{j} P_j,\cr
\tilde m(z)=&{1\over \mu^2}\sum_{k=1}^nt_k\sum_{j=1}^k P_jz^{k-j},
}}
where we suppressed the flavor indices and $V(z)$ and the $m_k$'s
are the electric ones in \elth.

We introduced the degree $n-1$ polynomial $\tm(P,z)$, which can be conveniently
cast in the form\foot{We will always understand a factor ${1\over 2\pi i}$ in the
measure of the contour integrals.}
\eqn\pizzi{
\tilde m(z)=
{1\over\mu^2}\oint_A d\zeta{V'(\zeta)-V'(z)\over\zeta-z}P(\zeta),
}
where $A$ is a contour that sorrounds all the roots of $V\,'(z)$.
We introduced also a meromorphic function that collects for the
gauge singlets
\eqn\singi{
P(z)=P_1z^{-1}+\ldots+P_nz^{-n},
}
and note that the last term in the superpotential \magtre\ can be rewritten as
\eqn\magtres{
\oint_A m(z)P(z).
}
Moreover, by inverting \pizzi\ we find that the general expression for the singlets
is fixed by $\tm(z)$ to
\eqn\singlets{
P(z)=\mu^2 \left[{\tilde m(z)\over V'(z)}\right]_{-n},
}
meaning that we take the Laurent expansion up to ${\cal O}(z^{-n})$.
The equations of motion for the singlets are
\eqn\sina{\eqalign{
\sum_{i=j}^nt_i\tilde qY^{i-j}q&=-\mu^2 m_j,\qquad j=1,\ldots,l,\cr
\sum_{i=j}^nt_i\tilde qY^{i-j}q&=0,\qquad j=l+1,\ldots,n.
}}
Therefore, the classical chiral ring of this theory does not contain
the mesons $\tq Y^{j-1}q$, which are in fact replaced by the $n$ singlets
$P_j$. We already see here that the analysis of the chiral ring in this theory
will be slightly different than the usual electric one.
We will be still interested in the following chiral operators
\eqn\chiralma{\eqalign{
\tilde R(z)&= -{1 \over 32 \pi^2} \Tr {W_\alpha W^\alpha \over
z-Y},\cr
\tilde M_{f}^{\tf}(z) &= \tilde q^{\tilde f} {1\over z-Y} q_f,\cr
\tilde T(z)&= \Tr {1\over z-Y}.
}}
We already set to zero the magnetic $w_{\alpha}$ generator analogous
to the one in \chiralopf.

Now we want to look at the magnetic vacuum corresponding to the pseudoconfining
electric one in \convac. This phase is characterized by a vanishing
classical expectation value for the gauge singlets $P_j$, since they
represent to the electric mesons. We have to satisfy the singlet equations
of motion \sina, as well as the adjoint ones $V'(Y)=0$.
Consider at first the simple case in which only the last
flavor appears in the electric meson superpotential \elth, i.e.
$m(z)_f^{\tf}=m(z)_{N_f}^{N_f}$. Correspondingly, the right hand side
of the singlet equations of motion \sina\ has nonvanishing
entries only along these flavor directions. Let us denote
\eqn\billi{
b_{1}=\left(-{m_1\mu^2\over t_n}\right)^{1\over n+1},
}
which has the dimension of a mass, and introduce the following bra--ket notation
$$
|i\rangle\leftrightarrow i^\alpha=\delta^\alpha_i,
$$
where a ket corresponds to a field
in the fundamental representation of the gauge group and a bra to a field
in the antifundamental.
We introduce also the shift operator acting on the first $n$ entries
\eqn\shiff{\eqalign{
R_{n}|i\rangle&=\cases{|i-1\rangle\qquad i=2,\dots,n,\cr
0\qquad\qquad {\rm otherwise}.}
}}

In this notations, the classical expectation value for the adjoint can be
represented in block diagonal form
\eqn\pseuY{
Y=\diag(Y_{N_f},Y_{a_1},\ldots,Y_{a_n}),
}
where
\eqn\conadj{\eqalign{
Y_{N_f}=&|1\rangle\langle v_{N_f}|+b_1R_n,\cr
|v_{N_f}\rangle=&-\sum_{k=1}^{n-1}{t_{n-k}\over t_n}b_1^{1-k}|k\rangle.
}}
The first $n\times n$ block \conadj\ reads
\eqn\conadjo{
Y_{N_f}=\pmatrix{
-{t_{n-1}\over t_n}               &  b_1  &  0  &      &       &  &          \cr
-{t_{n-2}\over t_n}{1\over b_1}  &  0  & b_1  &  .   &       &  &        \cr
     .                                &     &  0  &  .   &    .  &          \cr
     .                                &     &     &  .   &   b_1   & 0          \cr
-{t_1\over t_n}{1\over b_1^{n-2}}               &  0  &  .  &  .   &   0    & b_1  \cr
    0                                &     &     &      &       . &    0        \cr
}}
The blocks $Y_{a_1},\ldots,Y_{a_n}$ in \pseuY\ correspond
to the electric pseudoconfining eigenvalues and
are given by
\eqn\yai{
Y_{a_i}=\diag(a_i,\ldots,a_i).
}
Each $Y_{a_i}$ block has rank $\bar N_i=N_f-N_i-1$. Note
that this is different from the usual
relation $\ov{N}_i=N_f-N_c$ that we have in each low energy Seiberg block
when the meson polynomial $m(z)$ is switched off. We can check
that with this solution the correct magnetic rank is reproduced. Since
$\sum_i N_i=N_c$, we have
\eqn\ranks{
\sum_{i=1}^n(N_f-N_i-1)+n=nN_f-N_c=\ov{N}_c.
}

The magnetic quarks are all vanishing except the last flavor
\eqn\confun{\eqalign{
|\tilde q^{N_f}\rangle=&b_1|1\rangle,\cr
|q_{N_f}\rangle=&\sum_{i=1}^{l+1}b_1^{i}{m_i\over m_1} |n+1-i\rangle,
}}
whose vevs
are along the first $n$ color directions, in order to sandwich the first
block $Y_{N_f}$ in the adjoint and satisfy the singlet equations
of motion.
Note that, in the simplest case in which $V'(z)=t_nz^n$ and $m(z)^f_{\tf}=m_{N_f}^{N_f}$, this solution
reduces to the usual KSS solution \KutaS\KSS. We can also write down the classical expressions
for the generators \chiralma\ for this particular solution. The resolvent $\tilde R(z)$
vanishes while
\eqn\clasmag{\eqalign{
\tilde M_{cl}(z)^{N_f}_{N_f}=&-\mu^2{m(z)\over V'(z)},\cr
\tilde T_{cl}(z)=&{d\over dz}\ln{V'(z)\over z}+{1\over z}+\sum_{i=1}^n{\bar N_i\over z-a_i},
}}
where $m(z)$ is the electric meson polynomial. Moreover, at large $z$ we find
$\tilde T_{cl}\sim \oN_c/z$ since $\sum_i\oN_i=\oN_c-n$. In the electric
pseudoconfining phase, the magnetic singlets vanish classically, thus
we find that in the classical chiral ring $\tilde m(z)=0$.

The low energy theory described by \conadj\ can be studied in two steps, following
the KSS procedure.
First, the $n\times n$ block \conadjo\ higgses the theory down to
$$
U(\oN_c)\to U(\oN_c-n),
$$
and  note that $\oN_c-n=n(N_f-1)-N_c$ as expected from the electric theory, where
we integrated out the last massive flavor.
At this stage, $\tilde q^{N_f}_\alpha$, $q_{N_f}^\alpha$,
$Y^\alpha_m$, $Y_\alpha^s$ for $\alpha=n+1,\ldots,\oN_c$ and
$m=2,\ldots, n$, $s=1,\ldots,n-1$ conspire to join $n$ massive vector superfields in the fundamental
representation of the low energy gauge group $U(\oN_c-n)$ with mass squared
$b_1^2$. But then as we decompose the adjoint
we find that the higgsed flavor gets replaced by a new flavor
$Y_1^\alpha$, $Y_\alpha^{n}$ for $\alpha=n+1,\ldots,\bar N_c$,
so the number of flavors does not decrease here.
Secondly, the superpotential for the adjoint generates a mass term for this new flavor.
Only the leading term $\Tr Y^{n+1}$ contributes
\eqn\massmag{
{t_n\over n+1}\Tr Y^{n+1}=t_n Y^\gamma_\alpha\langle Y^{n-1}\rangle^\alpha_\beta Y^\beta_\gamma=
t_n b_1^{n-1}Y^\gamma_1 Y^n_\gamma.
}
The number of flavors effectively decreases by one unit also in the magnetic theory.
The singlets $(P_j)^i_{N_f}$ and $(P_j)^{N_f}_i$, $i=1,\ldots,N_f$ become also massive.
Now we can set the massive fields to the solution of their equations of motion and integrate them out.
The effective superpotential at a scale below $b_1$ is $\Tr V(\hat{Y})+\hat{\tq}\tm(\hat Y)\hat q$,
where the hatted fields transform in the representation of the low energy
gauge group $U(\oN_c-n)$ and we are left with $N_f-1$ flavors.
The matching of the scales goes as follows
\eqn\magmat{
\tilde{\Lambda}^{2\oN_c-N_f}_{\oN_c, N_f}={m_1\mu^2\over t_n^2  }
\tilde{\Lambda}^{2(\oN_c-n)-(N_f-1)}_{\oN_c-n,N_f-1}   .
}
%In the electric theory we just integrated out the last flavor, getting
%a scale matching $m_1\Lambda^{2N_c-N_f}_{N_c,N_f}=\Lambda^{2N_c-(N_f-1)}_{N_c,N_f-1}$.
We can use the relation  between the scales \scalerel\ and the electric scale
matching and find
\eqn\match{
\Lambda^{2N_c-(N_f-1)}_{N_c,N_f-1}
\tilde{\Lambda}^{2(\bar N_c-n)-(N_f-1)}_{\bar N_c-n,N_f-1}=\left({\mu^2\over t_n^2}\right)^{N_f-1}.
}
If we keep flowing to energies below the $a_i$ of \yai\ we will find
the usual product of the magnetic theories dual to each electric SQCD
block.

This solution can be generalized to the case in which the electric
meson polynomial has nonvanishing entries
on different flavors. If also the one but last flavor appears in the
electric superpotential, i.e. $m(z)^{\tf}_f=m(z)^{N_f}_{N_f}+p(z)^{N_f-1}_{N_f-1}$ with
$\deg m(z)^{N_f}_{N_f}=l$ and $\deg p(z)^{N_f-1}_{N_f-1}=l'$,
then the new flavor contributes an additional $n\times n$ higgsed block
\eqn\twofl{
Y=\diag(Y_{N_f},Y_{N_f-1},Y_{a_1},\ldots,Y_{a_n}),
}
where the first block is always \conadj\ and the second block is similar but with the substitution
$b_1\to b'_{1}$. For what concerns the magnetic quarks, in addition to \confun\
also the one but last flavor is higgsed as follows
\eqn\confun{\eqalign{
|\tilde q^{N_f-1}\rangle=&b'_{1}|n+1\rangle,\cr
|q_{N_f-1}\rangle=&\sum_{i=1}^l{p_i\over p_1} (b'_1)^i|2n+1-i\rangle.
}}
The magnetic gauge group is now higgsed down to $U(\oN_c-2n)$ and in each low energy
Seiberg block we have the correspondence $\oN_i=N_f-N_i-2$.

This classical analysis can be pushed further until we hit the following bound on the
number of massless quarks
\eqn\stabi{
N_f\geq {N_c\over n}.
}
Suppose in fact that in the meson polynomial $m(z)$ there appear $N_f-N_c/n$
flavors so that
we saturate the bound \stabi. Then the magnetic gauge group would be
completely higgsed and we will see no low energy SQCD blocks. The solution to this problem
is that as the magnetic gauge group is completely higgsed, a new
superpotential is triggered by instantons in the broken gauge group and the singlet equations
of motion get modified.

\subsec{The Electric Stability Bound and Magnetic Instantons}

Let us briefly describe the stability bound on the electric theory \KutaS.
We have seen that a superpotential \elth\ drives the theory to a product
of low energy decoupled SQCD, breaking the gauge group down to
$\prod_{i=1}^n U(N_i)$. Consider each $U(N_i)$ SQCD block separately: it is well
known that this gauge theory admits a stable vacuum iff the number of flavors is larger
than the number of colors \ADS, i.e. $N_f\geq N_i\,\,\forall i$.
Therefore the original theory admits a stable vacuum iff the bound \stabi\ is
satisfied.\foot{The theory is always stable if all the flavors are massive,
which is the case we will consider when solving the quantum theory.}

When we completely break the magnetic gauge group, the weak coupling analysis
we carried out
is no longer valid due to the presence of instantons. A well known example is
$SU(N_c)$ SQCD with $N_f=N_c+2$ flavors
and its magnetic dual with gauge group $SU(N_f-N_c)=SU(2)$
\SeibergPQ.
If we add a mass term for the last electric flavor,
the magnetic gauge group gets completely higgsed, so that
instantons in the broken $SU(2)$ generate a superpotential term.
By passing to the electric variables, one can see that the sum
of the magnetic tree level and instanton superpotentials reproduces
the usual nonperturbative superpotential of SQCD with $N_f=N_c+1$.

We would like to generalize this issue to our case of adjoint SQCD
and check whether we can generate an instanton term in the magnetic
superpotential when approaching the stability bound.
We consider the case in which $\oN_c=n+1$, i.e. we have
$N_f=N_c/n+1+1/n$ flavors. At this point, we are just above the
bound \stabi\ and our classical analysis still makes sense.
We further specialize to $n=2$ and take the electric
deformation to be just $t_2\Tr X^3$.\foot{We
consider the $\oN_c=n+1$ rather than the $\oN_c=n$ case because in the latter
the magnetic deformation $\Tr Y^{n+1}$ is trivial in the classical
chiral ring and the analysis of the instantons is more involved due the presence
of additional flat directions.} Note that we do not
include a mass term for the adjoint. We further add
a mass term for the last flavor.
Our electric superpotential reads
\eqn\elettra{
W_{el}={t_2\over3}\Tr X^3+m\tQ_{N_f}Q^{N_f}.
}
The magnetic theory is a $U(3)$ gauge theory defined by the superpotential
\eqn\napis{
W_{mag}=-{t_2\over3}\Tr Y^{3}+{t_2\over\mu^2}(M_1\tq Yq+M_2\tq q)+m(M_1)^{N_f}_{N_f}.
}
The classical solution \conadj\ still applies and
the magnetic gauge group is higgsed down to $U(1)$.
Since the low energy dynamics is abelian, we may expect instanton effects in the broken
gauge group.
One can perform a standard analysis of the zero modes in the instanton
't Hooft vertex $\tilde{\L}^{2\oN_c-N_f}\lambda^{2\oN_c}\psi_Y^{2\oN_c}\psi_{\tq}^{N_f}\psi_q^{N_f}$,
where $\psi_\Phi$ denotes the second component of the chiral superfield $\Phi$.
By using the interactions in the tree level action, such as the scalar--fermion--gaugino
D--term vertex as well as the superpotential couplings in \napis, one can read out from
this vertex the following
contribution to the superpotential
\eqn\nacchio{
W_{inst}={t_2^{2N_f+3}\over m^2\mu^{4N_f}}(\L^{6-N_f})^2\det\hat{M}_2 (\hat{M}_1\cof\hat{M}_2),
}
where $\cof{M}\equiv M^{-1}\det{M}$ and the hatted fields transform
in the $SU(N_f-1)$ low energy flavor symmetry group \Csaki. Note
that this is the contribution by a two--instanton. As explained in \Klein,
this is due to the absence of the mass coupling for the adjoint, that would
have been an overall factor in the one--instanton term.

We see that, when hitting the bound
\stabi, the classical solution \conadj\ is no
longer valid, due to the presence of the instanton term that
couples the singlets.
We can also translate this superpotential to the electric
variables by using the scale matching relation \scalerel\ and the electric
low energy scale $\L_{low}^{2N_c-(N_f-1)}=m\L^{2N_c-N_f}$,
obtaining the electric superpotential
\eqn\elebubbo{
W_{nonpert}={\det M_2( M_1\cof M_2)\over t_2^{2N_f}(\Lambda^{2N_c-N_f})^2},
}
In this expression we dropped the hats and the subscript on the scale. It is to be understood
as the superpotential of a theory with $N_c$ colors and $N_f=(N_c+1)/2$ flavors.
The magnetic instanton superpotential is seen on the electric side as a nonperturbative
superpotential arising from strong coupling effects \Csaki, in a very similar way to
ordinary SQCD with $N_f=N_c+1$ flavors.

\subsec{The Chiral Ring}

In this section we will find the chiral ring of the quantum theory
by solving the generalized Konishi anomaly equations \CDSW\CSW.
In appendix A we quote the results we need about DV to set the notations, while
for a basic review and a guide to the vast literature we refer to \Ferretti.
As a first step,
we will consider the
case in which the meson superpotential is just a mass term for
all the flavors, with no Yukawa--type interactions between quarks and adjoint. While
the solution of the electric theory is standard, the anomaly equations in the magnetic
theory are somewhat different, due to the presence of the gauge singlets. This massive case
is useful to illustrate the general procedure without worrying about the
rich analytic structure of the generators of the chiral ring, that we will
encounter later.

\vskip0.2cm
\centerline{\it The Electric Theory}
\vskip0.2cm

We will focus on the case in which the electric meson superpotential
is just a mass term
\eqn\treesupa{
W_{el}=\Tr \,V(X) + \tilde Q_{\tilde f}
m^{\tilde f}_f  Q^f,
}
where $m$ is a diagonal matrix. If the second derivatives of $V(z)$ at the
saddle points are nonvanishing, all the fields will be massive and
it makes sense to use the effective action as a function of the glueball superfield $S$.
We will be interested in the chiral operators \chiralopf.

The solution of the anomaly equation for the resolvent $R(z)$ gives
\eqn\solveR{
2R(z)= V'(z) -\sqrt{V'(z)^2 +f(z)},
}
where $f(z)$ is a $n-1$ degree polynomial $f(z)=f_1+...+f_n z^{n-1}$.
This defines the curve of the electric theory to be the hyperelliptic
Riemann surface $y^2=V'(z)^2+f(z)$.

Since the meson polynomial $m(z)$ is just constant, the anomaly equation for
the matrix $M(z)$ reduces to the following simple form $[M(z)m]_-=R(z)$,
where we suppressed flavor indices. The solution is
\eqn\anoai{
M(z)=R(z)m^{-1},
}
$M$ being a diagonal matrix.

The anomaly equation for $T(z)$ is
\eqn\anoti{
[y(z)T(z)]_-+[\tr \,m'(z)M(z)]_-=0,
}
but since $m(z)=\const$ the last term drops.
The solution is
\eqn\solveT{
T(z)={c(z)\over \sqrt{V'(z)^2 +f(z)}},
}
where $c(z)$ is another $n-1$ degree polynomial $c(z)=c_1+...+c_n z^{n-1}$.
Since $m$ is not $z$-dependent, in
the electric theory the fundamentals do not influence the solution for $T$ .

The parameters $f_j,c_j$ are related
to the glueballs $S_i$ of the low energy SQCD blocks
and the ranks $N_i$ of their gauge groups as follows
\eqn\contin{\eqalign{
S_i &=\oint_{A_i} R(z) dz,\cr
N_i&=\oint_{A_i} T(z) dz,}}
where $A_i$ is classically a contour around $a_i$. At the quantum level,
each stationary point $a_i$ opens up into a branch cut for $R(z)$ and the contour
$A_i$ actually encircles the two branch points.
One can get exact formulae for the total glueball $S=\sum_i S_i$ and the rank of the
high energy gauge group $N_c=\sum_i N_i$
by looking at the $1/z$ terms in \solveR\ and \solveT, since choosing
a contour $A$ around all the branch points is equivalent to closing it around $\infty$.
In this way we can fix the first coefficient of the polynomials $c(z)$ and $f(z)$
\eqn\solveS{
S=-{f_n\over4 t_n},\qquad N_c={c_n\over t_n}.}

We calculate now the relevant relations in the chiral ring.
We can extract from \anoai\ the mesons operators by
\eqn\comme{
\tilde QX^{j-1}Q=\oint_A z^{j-1}M(z),
}
where the contour $A$ encircles all the branch points of the resolvent $R(z)$,
obtaining
\eqn\elmesono{
\tilde Q X^{j-1}Q =-\sum_{i=1}^n{a_i^{j-1}f(a_i)\over 4mV''(a_i)},
}
for $j=1,\ldots,n$, coming from the negative power expansion of the
first term in the semiclassical expansion of the resolvent
\eqn\semifi{
R(z)=-{f(z)\over 4 V'(z)}+{\cal O}\left({f(z)^2\over V'(z)^3}\right),
}
In particular we find the usual Konishi anomaly
\eqn\fundo{
\tilde Q_f Q^f = {N_f S \over m },
}
where we used \solveS. Higher meson operators
receive additional contributions from the semiclassical expansion.
The single trace of the adjoint $X$ can be obtained as the coefficients  of
inverse powers of $z$ in the expansion of $T(z)$ at large $z$
\eqn\firing{
\Tr X^j =\oint_{A}z^j T(z) dz,
}
where $A$ circles all the the branch points of the resolvent $R(z)$.
Expanding $T(z)$ we get
\eqn\asyT{
T(z)={c(z)\over  V'(z)} -{c_n f_n \over 2 t_n^3 z^{n+2}}+\ldots
}
and we can extract the chiral operators
\eqn\assyT{
\Tr X^j =\sum_{i=1}^n {c(a_i) a_i^j\over  V''(a_i)}+\delta^j_{n+1}{2 N_c S\over t_n},
}
for $j=1,\ldots,n+1$. Clearly the equation $V'(X)=0$ is obeyed in the chiral ring, but relations
obtained by multiplying it with $X$ get quantum corrections.

\vskip0.2cm
\centerline{\it The Magnetic Theory}
\vskip0.2cm

The magnetic theory corresponding to \treesupa\ has a tree level
superpotential
\eqn\treesupma{W_{mag}=\Tr \oV(Y) + \tilde q^{\tilde f}
\tm_{\tilde f}^f (Y) q_f + \om \tr(P_1).
}
Note that the quantities appearing in \treesupma\ are the magnetic ones, as explained at the beginning
of this section. In particular we have that
\eqn\pizzia{
\tm(z)=-{1\over\mu^2}\oint_{\tilde A}{\oV\,'(\zeta)-\oV\,'(z)\over \zeta-z}P(\zeta),
}
and, inverting this, we find the gauge singlets
\eqn\singla{
P(z)=-\mu^2\left[{\tm(z)\over \oV\,(z)}\right]_{-n},
}
We are ready to use now the anomaly equations.
The form of $\tilde R(z)$, which is independent on the fundamentals,
will be the same as for the electric theory
\eqn\solver{
2\tilde R(z)=\oV\,'(z)-\sqrt{\oV\,(z)^2+\bar f(z)},
}
where the quantum deformation $\bar f(z)$ is a degree $n-1$ polynomial.
Since we will see that
the quantum deformations on both sides are equivalent
under the offshell duality map, we will conclude that the magnetic theory
has the same curve of the electric one.

In addition to the usual anomaly equations, that we encountered in the
electric theory, there are new ones following from variations of the gauge singlets $P$'s.
Since $P$ is not coupled to the gauge fields, these are just its equations of motion.
For the special case we are studying, after rearranging the equations, \sina\
reduce to
\eqn\qeq{\eqalign{
\ot_n\tilde q Y^{n-1} q &= \om \mu^2,\cr
\tilde q Y^{j-1} q &=0, \qquad  j=1,...n-1.}}
On the other hand, the role of the electric meson polynomial is played now
by $\tm(z)$. The anomaly equation for the meson generator is then
\eqn\anomme{
[\tilde M(z)\tm(z)]_-=\tilde R(z).
}
Its generic solution is
\eqn\anoai{
\tilde M(z)=\tilde R(z)\tm^{-1}(z) +r(z) \tm^{-1}(z),
}
in our case all the matrices being diagonal. The crucial piece of information about
the magnetic theory is the quantum expression of $\tm(z)$, which
contains the gauge singlets and fixes the analytic properties of the meson
generator. The way in which \anoai\
is supposed to be used is the following
\item{1.} We fix the polynomial $r(z)$ such that
there are no additional singularities in \anoai\ arriving from the zeroes of $\tm(z)$.
\item{2.} We fix the polynomial $\tm(z)$ imposing that the mesons $\tilde q X^{j-1} q$
extracted from \anoai\ fulfill the singlet equations of motion
\qeq. In this way we fix also $P(z)$.

\noindent The unique solution to these
requirements is
\eqn\solP{
r(z)=0,
}
and
\eqn\solPP{
 \tm(z)= -{ \ov{f}(z)\over 4 \om \mu^2}.
 }
Since $\tilde m(z)$ is proportional to $\ov{f}(z)$ and the the resolvent $\tilde R(z)$ vanishes at the
zeroes of $\ov{f}(z)$, we see that $\tilde m^{-1}(z)$ does not give additional singularities in
\anoai. The analytic structure of $\tilde M(z)$ in this case turns out to be very simple, while
the singlets are
\eqn\solP{
P(z)={1\over 4 \om}\left[{\ov{f}(z)\over \oV\,'(z)}\right ]_{-n},
}
where the expansion in inverse powers
of $z$ is understood to stop at $z^{-n}$. Comparing with \elmesono\ we see that the matching
\eqn\match{
P_j= \tilde Q X^{j-1} Q,
}
for $j=1,\ldots,n$, is implied for a sign choice which will be discussed later.
Of course we could go backwards and requiring \match\
prove the form of the Kutasov kernel ${\oV\,'(\zeta)-\oV\,'(z)\over \zeta-z}$
which determines the form of the fundamental magnetic superpotential.
We can extract the expectation values of the magnetic singlets out of \solP
\eqn\singlve{
P_j={1\over 4\om}\sum_{i=1}^n{\of(\oa_i)\oa_i^{j-1}\over \oV\,''(\oa_i)}=
-\sum_{i=1}^n{\oa_i^{j-1}\bar{S}_i\over \om},
}
where we used the definition of the glueballs in \contin.

We can now calculate $\tilde T(z)$. Its anomaly equation is
\eqn\magti{
[\tilde y(z)\tilde T(z)]_-
+\tr[\tm'(z)\tilde M(z)]_-=0.
}
The solution here, as opposed to \solveT, depends also on the fundamentals
\eqn\magT{
\tilde T(z)={1\over\tilde y(z)}[ -\tm'(z)\tilde M(z)+\ov{c}(z)]
}
with $\tm'(z)$, $\tilde M(z)$ given by \solP\ and \anoai. Since $\ov{c}(z)$ is a
polynomial of degree $n-1$, while $\tm'(z)\tilde M(z)$  starts with $ z^{-2}$,
the contribution of the fundamentals will start only from the power
$z^{-n-2}$. Recalling that $\tm$ is a diagonal matrix, we
expand \magT\ at large $z$
\eqn\magTla{
\tilde T(z)={\ov{c}(z)\over \oV\,'(z)}-{\ov{c}_n \of_n\over 2\ot_n^2 z^{n+2}}
+N_f{\of_n\over4\ot_n^2}{1\over z^{n+1}}{\of'(z)\over \of(z)},
}
and the chiral ring is
\eqn\magchiad{
\Tr Y^j=\sum_{i=1}^n{\oa_i^j\ov{c}(\oa_i)\over \oV\,''(\oa_i)},
}
for $j=0,\ldots,n$. The first operator which will receive a contribution from the last two terms
in \magTla\ will be $\Tr Y^{n+1}$ which in the magnetic theory becomes\foot{
We used the fact that $\oint f'/f=\# {\rm \ zeroes\ of}\,f$.}
\eqn\asyTnmag{
\Tr Y^{n+1}=\sum_{i=1}^n {\ov{c}(\oa_i) \oa_i^{n+1}\over  \oV\,''(\oa_i)} +
{2 \oN_c \bar{S}\over \ot_n} - N_f {\bar{S}(n-1)  \over \ot_n}.
}

We would like to stress again a basic property of the solution \solPP. Since it is
proportional to $\of(z)$, the meson generator $\tilde M(z)$ and also $\tilde T(z)$ have a
very simple analytic structure in both sheets, as opposed to the generic cases
we will solve below. Because of this fact, we will be able to see that the
electric--magnetic duality map here works exactly offshell.

\subsec{The Effective Actions}

Once that we have solved the chiral ring,
we can determine the superpotential part of the low energy effective action by integrating
the derivatives with respect to the parameters appearing in the lagrangian, which are the
expectation values of the chiral operators we just computed above. This offshell effective action
will be valid at energies above the glueball mass, that sets the scale of mass gap.

\vskip0.2cm
\centerline{\it The Electric Theory}
\vskip0.2cm

The electric couplings are
$m$ and $t_1,\ldots,t_n$.  It is covenient to use as independent parameters $t_n$ and
 $\hat t_j={ t_j \over t_n} $ for $ j=1,...,n-1 $. The parameters
$\hat t_j$ are homogenous polynomials in $a_i$.
The derivatives of the effective action are
\eqn\phiexp{
 {\partial W_{eff} \over \partial \hat t_j}=
t_n {1\over j+1} \langle \Tr X^{j+1} \rangle = -{1\over j+1}\sum_{i=1}^n {t_n c(a_i) a_i^{j+1}\over 4 V''(a_i)},}
 for $ j=1,...,n-1 $  and
\eqn\phiexpn{
 {\partial W_{eff} \over \partial  t_n}=
 {1\over n+1} \langle \Tr X^{n+1} \rangle = -{1 \over n+1}\sum_{i=1}^n { c(a_i) a_i^{n+1}\over 4 V''(a_i)} +
{1 \over n+1} {2 N_c S\over t_n},}
\eqn\mexp{
\langle \tilde Q_f Q^f\rangle=
 {\partial W_{eff} \over \partial m }={N_f S\over m}.}
Since we are looking for the offshell effective action, these equations are supposed to be integrated
at fixed $S_i$, $N_i$. Now observe that \mexp\ and the second term in \phiexpn\ satisfy the integrability
condition by themselves. Therefore we can integrate them separately and
there is a solution ${\cal W}_{eff}$ without them. The general effective action we obtain by
\phiexp, \phiexpn\ and \mexp\ is
\eqn\gen{
W_{eff}= {\cal W}_{eff}  + {2 N_c S \over n+1} \log t_n +N_f S \log m
+ [t_j ,m- {\rm independent \ terms}]
}

Let us consider the coupling independent terms. There are two contributions, the first
is the one--loop exact renormalization of the gauge field kinetic term $(2N_c-N_f)S\log\Lambda$, that
contains the dynamically generated scale $\Lambda$ through the running gauge coupling
constant. Then we have a Veneziano--Yankielowicz type superpotential $bS(\log S-1)$. One can fix
the numerical coefficient $b$ by requiring that the effective action is $U(1)_R$ invariant.
Since the $R$--current and the dilatation current lie in the same ${\cal N}=1$ supermultiplet,
this is the same as fixing them by dimensional analysis. By the usual localization trick,
we promote the couplings to background chiral superfields so that we can assign them a charge.
The dimensions $\Delta$ of the various fields are
\eqn\eldim{
\matrix{ & \Delta \cr
S & 3  \cr
t_j & 2-j \cr
m   & 1\cr
\Lambda^{2N_c-N_f}& 2N_c-N_f}
}
so that we find $b=-2N_c/(n+1)$. Since ${\cal W}_{eff}$ is invariant
by itself we get the effective superpotential
\eqn\elfin{
W_{eff}={\cal W}_{eff} +S \log{ \Lambda ^{2N_c-
N_f} t_n^{2N_c \over n+1}m^{N_f} \over S^{2 N_c \over n+1}}+{2N_c\over  n+1}S.
}

We will now turn to the evaluation of the term ${\cal W}_{eff}$. It is most
convenient to parameterize the degree $n-1$ polynomial $c(z)$ in the following way
\eqn\defc{ c(z)=V'(z)\sum_{i=1}^n{h_i\over z-a_i}.
}
where $N_c=\sum_{i=1}^{n} h_i$.
The $n$ coefficients $h_i$ are fixed by the contour integral
\eqn\bicon{
h_i=\oint_{A_i}{c(z)\over V'(z)},
}
so that classically we have just $h_i=N_i$. Using this parametrization
we can rewrite the relevant part of \phiexp\ and \phiexpn\ as
\eqn\as{
{\partial{\cal W}_{eff}\over \partial t_j }=\sum_{i=1}^{n} h_i a_i^j.
}
In particular, we see that $\Tr X^j=\sum_i h_ia_i^j$ for $j=1,\ldots,n$,
while $\Tr X^{n+1}$ contains in addition the last term in \assyT.
The coefficients $h_i$ depend on $t_j$, $S_i$ and $N_k$, as we can see
from \bicon. It is convenient to use in place of the glueballs $S_i$ the
new variable
\eqn\sbrabra{
y=\sum_{i=1}^n\log S_i,
}
and $n-1$ independent ratios of glueballs, e.g. ${S_1\over S_n},\ldots,{S_{n-1}\over S_n}$.
Introduce now the following functions
\eqn\dideffi{
d_i=h_i-e^y\int_{-\infty}^y dy\,' e^{-y'}{\partial h_i\over \partial y'}.
}
We claim that integrating \as\ we find
\eqn\tildedabbiu{
{\cal W}_{eff}=\sum_{i=1}^n d_iV(a_i).
}
In the Appendix B we prove, up to an assumption of integrability,
that indeed differentiating \tildedabbiu\ one recovers \as.

Putting everything together, the effective superpotential of the electric theory
is
\eqn\effeel{
W_{eff}=\sum_{i=1}^n d_iV(a_i)+S \log { \Lambda ^{2N_c-
N_f} t_n^{2N_c \over n+1}m^{N_f} \over S^{2 N_c \over n+1}}+{2N_c\over n+1}S.
}

\vskip0.2cm
\centerline{\it The Magnetic Theory}
\vskip0.2cm

We can follow again the same procedure of integrating the expectation values
with respect to the parameters, but now we have a new coupling $\mu$ and the
derivative with respect to $\ot_j$ gets a contribution also from the magnetic
fundamentals and singlets
\eqn\prapra{
{\partial \ov{W}_{eff}\over\partial \ot_j}={1\over j+1}\langle\Tr Y^{j+1}\rangle+
{1\over\mu^2}\sum_{i=1}^j\langle P_i \tilde qY^{j-i}q\rangle.}
for $j=1,\ldots,n$. Then we have
\eqn\prappo{
\mu^2{\partial \ov{W}_{eff}\over\partial\mu^2}=\om\langle \tr P_1\rangle,\qquad
{\partial \ov{W}_{eff}\over\partial \om}= \langle\tr P_1\rangle,
}
where we used the fact that the expectation values of gauge invariant chiral operators
factorize and the singlet equations of motion \sina,
which are exact in the quantum theory. We can substitute the expectation values \singlve, \magchiad,
and \asyTnmag\ into \prapra, obtaining
\eqn\expll{
{\partial \ov{W}_{eff} \over \partial  \ot_j}=
{1 \over j+1}\sum_{i=1}^n { \ov{c}(\oa_i) \oa_i^{j+1}\over  \oV\,''(\oa_i)}
+{\delta^j_n \over n+1} {2 (\oN_c+N_f) \bar{S}\over \ot_n},
}
for $j=1,\ldots,n$ and into \prappo\
\eqn\expa{
\mu^2{\partial \ov{W}_{eff}\over\partial\mu^2}=-N_fS,\qquad
{\partial \ov{W}_{eff}\over\partial \om}=-N_f{S\over \om}.
}
The first term in \expll\ is analogous to the corresponding electric one in \phiexp.
We assume, as in that case, that it satisfies the integrability condition by itself and integrate
it to obtain $\ov{{\cal W}}_{eff}$. This is formally equal to \tildedabbiu\ but
with magnetic quantities instead.
On the other hand, also \expa\ and the second term in \expll\
satisfy the integrability condition, so that we find
\eqn\magfiri{
\ov{W}_{eff}=\ov{{\cal W}}_{eff}+{2(\oN_c+N_f)\over n+1}\bar{S}\log \ot_n-2N_f\bar{S}\log \mu
-N_f\bar{S}\log \om+
[\ot_j ,\om,\mu- {\rm indep. \ terms}]
}
Then we need to add the magnetic one--loop
renormalization of the gauge fields $(2\oN_c-N_f)\bar S\log\tilde{\Lambda}$ and the Veneziano--Yankielowicz type
superpotential $\bar{b}\bar S(\log \bar S-1)$. Again we fix the coefficient $\bar{b}$ requiring
$U(1)_R$ invariance, as we did for the electric case, and get $\bar{b}=2(nN_f-\oN_c)/(n+1)$.
Putting everything together we obtain the magnetic effective action
\eqn\effemag{
\ov{W}_{eff}=\sum_{i=1}^n \bar d_i \oV(\oa_i)+\bar S\log {\tilde{\Lambda}^{2\oN_c-N_f}\ot_n^{2(\oN_c+N_f)\over n+1}
\bar S^{2(nN_f-\oN_c)\over n+1}\over \om^{N_f}\mu^{2N_f}}-{2(nN_f-\oN_c)\over n+1}\bar S. }

\subsec{The Offshell Duality Map}

At this point we will look for the duality map between the electric and magnetic operators
in the chiral ring. As we discussed in the introduction,
in this case the duality holds exactly offshell. First we will
consider the match of the meson operators and then the effective actions.

The gauge singlets equations of motion \sina\ are exact in the chiral ring of the
magnetic quantum theory. They tell
us that the magnetic meson operators $\tq Y^{j-1}q$ are trivial.
They are replaced by the gauge
singlets, which represent the electric mesons
through a Legendre transform, as it is clear from the expression of
$\tm(z)$ in the magnetic tree level
superpotential \magtre. Therefore we should match directly the electric mesons
with the corresponding magnetic gauge singlets through the relation
\eqn\matcha{
P_j=\tilde QX^{j-1}Q,
}
independently on the other relations between the gauge groups. Comparing
the two expressions \singlve\ and \elmesono\
$$
\tilde Q X^{j-1}Q =\sum_{i=1}^n{a_i^{j-1}S_i\over m},\qquad
P_j=-\sum_{i=1}^n{\bar{a_i}^{j-1}\bar{S_i}\over \om},
$$
for $j=1,\ldots,n$, we get the relations
\eqn\glurel{
S_i=-\bar{S_i},\qquad m=\om,
}
while the roots of the electric and magnetic polynomials for the adjoint coincide
$a_i=\bar a_i$, i.e. the electric polynomial $V'(z)$ and the magnetic
one $\ov{V}\,'(z)$ are identified up to a minus sign.
Let us recall the definition \contin\ of the glueballs in terms of
the resolvent
$$
S_i=-{f(a_i)\over 4V''(a_i)},
$$
which holds both for the electric and magnetic theories with the respective
quantities. The relation \glurel\ then fixes the the duality map as
\eqn\polrel{
f(z)=\bar{f}(z), \qquad V'(z)=-\ov{V}\,'(z).
}
This last relation,
in particular, tells us that electric and magnetic
theories have the same curve
\eqn\curve{
y^2=V'(z)^2+f(z).
}

Now let us consider the electric and the magnetic effective actions \effeel\ and \effemag.
By comparing their second and third terms we get again the match between
the glueballs and the mass terms \glurel\ and $t_n=-\ot_n$, which fixes
the ambiguity in the sign choice of \polrel, together
with
the scale matching relation
\eqn\scalematch{
\Lambda^{2N_c-N_f}\tilde{\Lambda}^{2\oN_c-N_f}= t_n^{-2N_f}\mu^{2N_f},
}
and the usual relation between the electric and magnetic gauge groups
$\oN_c=nN_f-N_c$.
The scale matching \scalematch\ is consistent with the
fact that $\log \Lambda^{2N_c-N_f}$ and $\log \tilde{\Lambda}^{2\oN_c-N_f}$
are the sources for the respective electric and magnetic total glueballs and that
we found $\bar S=-S$. Let us analyze in more detail the relation between
the gauge groups.
The rank of the electric and magnetic gauge groups
fixes the pole at infinity of $T(z)$
\eqn\ninti{
N_c=\oint_A T(z),
}
where $A$ is the large contour, and analogously for the magnetic theory.
The matching $\oN_c=nN_f-N_c$ translates into
the following relation
\eqn\matti{
\oint_{\tilde A} \tilde{T}(z)=N_f\oint_A {V'(z)\over V''(z)}-\oint_A T(z).
}
We evaluate the contour integrals by expanding \solveT\ and \magT\ at large $z$ and get
\eqn\matci{
\ov{c}(z)=c(z)-N_fV''(z).
}
By the definition \bicon\ of the coefficients $h_i$ we find that
$\ov{h}_i=N_f-h_i$ or equivalently $\ov{d}_i=N_f-d_i$, which fixes the map between
the operators
\eqn\adjomap{
\Tr Y^j=-\Tr X^j+N_f\sum_{i=1}^na_i^j,
}
for $j=1,\ldots,n$, in agreement with the KSS results \KSS.
The match between the electric and magnetic ${\cal W}_{eff}$ using the relation $\bar d_i=N_f-d_i$
shows that the magnetic effective superpotential contains an additional
$Y$--independent term, which in this case is just $N_f\sum_i V(a_i)$.

The classical limit of the coefficients $h_i$ is $N_i$, the rank of each
low energy SQCD block. Thus we recover the usual matching relation
$\oN_i=N_f-N_i$, which is somewhat different from the one we found in
our classical analysis of \ranks, which anyway
was only valid in the case where the number of massive
electric quarks is less than $N_f-N_c/n$, because of the stability bound.
The higgsed blocks in the magnetic adjoint we found in \pseuY\ and \twofl,
that were responsible for the singularities in \clasmag, are an artifact of the
classical theory. When we pass to the full quantum theory, in this pseudoconfining
case all the classical singularities of $\tilde M(z)$ are smoothed out, and this is the reason
why we get back the usual Seiberg duality map $\oN_i=N_f-N_i$ for the rank of the gauge groups
of the low energy SQCD blocks.

\subsec{The Generic Pseudoconfining Case}

In the last section we saw that, in the case of massive quarks without Yukawa couplings, duality
works offshell, that is at the level of the dynamical effective actions.
We will consider in this section the most generic pseudoconfining case, where
in addition to the mass terms for the quarks we allow for a generic $z$--dependent
meson polynomial. As a consequence, the analytic properties
of the various resolvents in the quantum chiral ring get more involved and
in the end the match between electric and magnetic quantities will not
hold anymore exactly offshell, but we expect it to hold only onshell.
We will not compute the effective action, as we did above, but we will
match the electric mesons with the magnetic singlets
and find a map that reproduces the Konishi anomaly in each low energy
SQCD block as a classical equation in the magnetic theory.

\vskip0.2cm
\centerline{\it The Electric Theory}
\vskip0.2cm

Let us consider the electric theory with a generic yet diagonal
meson polynomial
\eqn\elthi{\eqalign{
W_{el}=&\Tr V(X)+\tilde Q^{\tilde f}m(X)_f^{\tilde f} Q_f,\cr
m(z)_f^{\tilde f}=&\sum_{i=1}^{l+1}m_iz^{i-1}\delta_f^{\tilde f}
}}
We denote the roots of $m(z)$ as $x_k$, for $k=1,\ldots,l$. The degree
of the polynomial $m(z)$ is at most $n-1$ and its
constant term $m_1$ must be nonzero for all the flavors in order for the theory to be massive.
The classical pseudoconfining vacuum is \convac, while the generators \chiralopf\ of the
classical chiral ring all vanish except
\eqn\claschiel{
T_{cl}(z)=\sum_{i=1}^n{N_i\over z-a_i}.
}
This phase is characterized by a vanishing classical expection
value for the fundamentals.

Let us consider the generalized Konishi anomaly equations.
The resolvent $R(z)$ is still given by \solveR. The story is different for $M(z)$,
the generator of the mesons. When solving its anomaly equation, we
have to cancel the additional singularities coming
from the zeroes of $m(z)$. We have to specify the boundary conditions
coming from our choice of the vacuum. In this pseudoconfining case, $M(z)$
is regular in the first sheet (up to the residue at infinity).
Implementing these boundary conditions we find
\eqn\meselge{
M(z)={R(z)\over m(z)}-\sum_{k=1}^{l}{R(q_k)\over z-x_k}{1\over m'(x_k)}.
}
Let us extract the expectation value of the mesons. We can evaluate \comme\
by expanding semiclassically the resolvent in powers of $f(z)/V'(z)^2$ as in \semifi\ and find
\eqn\conelmes{
\tilde Q X^{j-1}Q=-\sum_{i=1}^n{a_i^{j-1}f(a_i)\over 4m(a_i)V''(a_i)}+\ldots,
}
where we showed only the leading approximation. Here
we see the crucial difference between the purely massive case \elmesono\ and this general case.
There, we took the semiclassical expansion and then we opened up
the contour $A$ to the big circle, throwing away all the higher terms in the expansion. Here,
we cannot open up the contour $A$ after taking the semiclassical expansion,
because in this process we would hit the additional poles at the zeroes of $m(z)$ for each
term in the expansion. Due to the richer analytic structure,
we are forced to keep in \conelmes\ all the semiclassical expansion.
We will see that a duality map exists at the first order
in this expansion.

A similar story carries on to the last anomaly equation \anoti,
whose solution with the classical limit \claschiel\ is
\eqn\contele{
T(z)=\sum_{k=1}^{l}{1\over 2(z-x_k)}-\sum_{k=1}^{l}{y(q_k)\over 2y(z)(z-x_k)}+{c(z)\over y(z)},
}
where
\eqn\cizeta{
c(z)=V'(z)\sum_{i=1}^n{h_i\over z-a_i}-{1\over2}\sum_{k=1}^{l}{V'(z)-V'(x_k)\over z-x_k},
}
is a degree $n-1$ polynomial whose leading coefficient is $c_n/t_n=N-l/2$. Note
that in this case the fundamentals do contribute to $T(z)$.
We have considered a convenient parametrization of \cizeta\ similar to the one in
\defc\ but now slightly modified to take into account the more
complicated analytic structure.
We still have $\sum_{i}h_i=N_c$. Since the roots
$x_k$ of the meson polynomial $m(z)$ are supposed to be very large
in the semiclassical limit, we see that the definition
of the coefficients $h_i$ is still \bicon, the last term in \cizeta\ not
contributing to the contour integral. Now we can integrate the generator on
the contour $A$ to obtain the expectation values. The first term in
\contele\ does not contribute because the $x_k$'s lie outside the contour and we obtain
\eqn\traggio{
\Tr X^j=\sum_{i=1}^nh_ia_i^j+
\sum_{k=1}^{l}R(x_k)\sum_{i=1}^n{a_i^j\over(a_i-x_k)V''(a_i)}+
\delta^j_{n+1}{2S(N-l/2)\over t_n}+\ldots,
}
for $j=1,\ldots,n+1$. By $\ldots$ we denote higher terms in the semiclassical expansion \semifi.

\vskip0.2cm
\centerline{\it The Magnetic Theory and the Match}
\vskip0.2cm

The magnetic theory corresponding to \elthi\ has the tree level superpotential
\eqn\maltre{
W_{mag}=\Tr \oV(Y)+\tq\tm(P,Y)q+\oint \om(z)P(z),
}
where we use the same notations as in \pizzia\ and $\om(z)$ corresponds to the electric
meson polynomial. This phase is characterized by a vanishing classical value
of the singlets and thus also of $\tm(z)$.

Let us solve the anomaly equations. The resolvent
$\tilde R(z)$ is still given by \solver. The anomaly equation for the generator of the magnetic mesons
is always \anomme, whose general solution is
\eqn\genemsgen{
\tilde M(z)=\tilde R(z)\tilde m^{-1}(z)+r(z)\tilde m^{-1}(z).
}
We recall that the polynomial $r(z)$ is fixed in order to cancel the additional singularities coming from
the zeroes of $\tilde m(z)$. Then $\tilde m(z)$ is fixed by imposing that the magnetic
singlet equations of motion are satisfied.
Denote the roots of the degree $n-1$ polynomial $\tilde m(z)$ as $e_k$, for $k=1,\ldots,n-1$.
In this case our boundary conditions are such that $\tilde M(z)$ is regular in the
first sheet at the zeroes $e_k$
\eqn\confer{
{r(z)\over \tilde m(z)}=-\sum_{k=1}^{n-1}{\tilde R(e_k)\over z-e_k}{1\over \tilde m'(e_k)}.
}
Note that in the previous case \solP\
there was no need to keep the polynomial $r(z)$, since
$\tilde m(z)$ was proportional to the quantum deformation $\of(z)$ of the resolvent.
In that case, no additional singularity was present. Now the story is quite different and
to find the result we should first rewrite the singlet equations of motion in a more
convenient way.
First note that, just as we can usually trade
the glueballs $S_i$ for the coefficients of the quantum deformation $f(z)$ \Cava, we can also trade
the $n$ singlets $P_l$ for the $n$ coefficients of the polynomial
$\tilde m(z)=\sum_{l=1}^n\tilde m_l z^{l-1}$, that are a linear combination thereof
\eqn\tildec{
\tilde m_l=-{1\over\mu^2}\sum_{k=l}^n \ot_k P_{k-l+1}.
}
Now we cast the superpotential in a suitable form to replace the $P(z)$ with the $\tilde m(z)$.
Recall that the singlets are fixed by $\tm(z)$ as in \singla.
By using \maltre, the relevant part of the superpotential we need is
\eqn\reltree{
\tilde q \tilde m(Y) q-\mu^2\oint_{\tilde A} {\om(z)\tilde m(z)\over \oV\,'(z)}.
}
Differentiating w.r.t. $\tilde m_l$ we get
\eqn\tildeom{
\tilde q Y^{l-1}q-\mu^2\oint_{\tilde A}z^{l-1}{\om(z)\over \oV\,'(z)}=0,
}
for $l=1,\ldots,n$, that we can also write as
\eqn\fineom{
\oint_{\tilde A}z^{l-1}\left[\tilde M(z)
-\mu^2{\om(z)\over \oV\,'(z)}\right]=0,
}
Note that while in the electric case the zeroes of $m(z)$ are very large
in the semiclassical regime, in the magnetic case it turns
out that the zeroes of $\tm(z)$ do lie inside the $\tilde A$ contour, as we will see explicitly
in section 4 for the cubic superpotential.
We can expand the resolvent semiclassically as in \semifi\
and only will the residue at the zeroes of $V'(z)$
contribute. Remember that the singlet equations of motion \fineom\ are supposed to
fix the unknown polynomial $\tm(z)$. Indeed the solution of \fineom\ at the first order
in the semiclassical expansion is
\eqn\conjea{
4\mu^2 \om(\oa_i)\tilde m(\oa_i)=-\of(\oa_i),
}
for $i=1,\ldots,n$, where $\oa_i$ are the roots of $\oV\,'(z)$ and the flavor indices are suppressed (note that
they are not summed over). Eq. \conjea\ consists of $n$ conditions
that account for the $n$ unknown coefficients $\tilde m_l$.

Some comments are in order. The classical limit of \conjea\ is well defined,
since both sides vanish (remember that classically the singlets vanish in this phase).
Now look at the meson generator $\tilde M(z)$ in \genemsgen\ with boundary conditions \confer.
At the quantum level it is regular in the first semiclassical sheet, while it
has $n-1$ poles on the second sheet. Nevertheless, when taking the classical limit, both
$\tilde R(z)$ and $\tm(z)$ vanish, but the result is a nonvanishing classical
value for $\tilde M(z)$, that reproduces our classical understanding of the theory being higgsed,
as explained in section 2.1. Here we see again the same issue discussed at the end
of section 2.5. The singularities of $\tilde M(z)$ on the first sheet are an
artifact of the classical theory in the pseudoconfining case:
they are smoothed out in the full quantum theory.

We can complete the analysis of the magnetic chiral ring by
solving the anomaly equation \magti\ for $\tilde T(z)$
where $\tilde m(z)$ and $\tilde M(z)$ are given by \genemsgen\ and
\conjea\ and get
\eqn\conadmag{
\tilde T(z)=\sum_{k=1}^{n-1}{1\over 2(z-e_k)}-\sum_{k=1}^{n-1}{\tilde y(e_k)\over2\tilde y(z)(z-e_k)}+
{\bar c(z)\over \tilde y(z)},
}
where we can choose the following parametrization for the degree $n-1$ polynomial $\bar c(z)$
\eqn\concizzi{
\bar c(z)=\oV\,'(z)\sum_{i=1}^n{\bar h_i\over z-\oa_i}
-{1\over2}\sum_{k=1}^{n-1}{\oV\,'(z)-\oV\,'(e_k)\over z-e_k},
}
where we can fix $\ov{c}_n/\ot_n=\oN_c-n+1$ and $\sum_i\bar h_i=\oN_c$.
By following the same procedure as in the electric case \traggio, we
can extract again the corresponding magnetic expectation values
\eqn\tramad{
\Tr Y^j=\sum_{i=1}^n\bar h_i\oa_i^j+
\sum_{k=1}^{n-1}\tilde R(e_k)\sum_{i=1}^n{\oa_i^j\over(\oa_i-e_k)\oV\,''(\oa_i)}+
\delta^j_{n+1}{\bar S(2\oN_c-n+1)\over \ot_n}+\ldots,
}
for $j=1,\ldots,n+1$. By $\ldots$ we denote the higher terms in the semiclassical
expansion.

Now we can check that our singlets $P(z)$ in \singlets\ match the electric
mesons \conelmes
\eqn\matchi{
\tilde Q X^{j-1}Q=-\sum_{i=1}^n{a_i^{j-1}f(a_i)\over 4m(a_i)V''(a_i)}+\ldots,
\qquad P_l=-\sum_{i=1}^n{\oa_i^{l-1} \tilde m(\oa_i)\mu^2\over
\ov{V}\,''(\oa_i)}.
}
At the leading approximation in the electric semiclassical expansion \semifi, the match is ensured
by the condition \conjea\ that solves the singlet equations of motion, provided
that the relation between electric and magnetic
polynomials and the quantum deformations is again
\eqn\matchv{\eqalign{
\ov{V}\,'(z)=&-V'(z),\qquad \om(z)=m(z),\qquad \ov{f}(z)=f(z),
}}
just the same we found in the massive case \polrel. Therefore electric and magnetic
theory still have the same curve \curve.
Note that this is equivalent to the simple relation between the glueballs
$$\bar{S}_i=-S_i.
$$
At this point we can rewrite the solution \conjea\ of the magnetic theory is terms
of the electric quantities, recalling that $\tm(z)$ in \pizzia\ reverses its sign
\eqn\conje{
4\mu^2 m(a_i)\tilde m(a_i)=f(a_i).
}

\vskip0.2cm
\centerline{\it The Konishi Anomaly}
\vskip0.2cm

Consider the low energy theory described by the vacuum \convac. It is a
product of decoupled SQCDs with $N_f$ flavors. We can look at the
physics of each separate $U(N_i)$ SQCD by integrating the resolvents around
the contour $A_i$, that encircles the branch points of the
resolvent appeared by the splitting of the $a_i$ root. In particular,
by \conelmes\ the mesons, even if classically vanishing,
at the quantum level satisfy the Konishi anomaly
\eqn\koniss{
\langle \tQ Q\rangle_i={S_i\over m(a_i)}=-{f(a_i)\over4m(a_i)V''(a_i)},
}
where we dropped the higher terms in the semiclassical expansion \semifi.

We can perform a similar analysis in the magnetic theory: the
electric mesons in \koniss\ correspond to the gauge singlets $P_1$.
The general expression for the singlets is given in \singla. The low
energy magnetic theory is a product of decoupled $n$ Seiberg blocks,
each one dual to a corresponding electric SQCD. The relation corresponding to
the Konishi anomaly \koniss\ in each low energy block is
\eqn\correp{
\langle P_1\rangle_i=-\mu^2{\tm(\oa_i)\over \oV\,''(\oa_i)},
}
and it matches the electric one due to the relations \conjea\ and \matchv.

\newsec{The Electric Higgs Phase}

In this section we will find the classical magnetic solution
when the electric theory is in the higgs vacuum and check its properties. Then we will solve the chiral
rings and look for the duality map. In the end we will
consider the analytic structure of the solution we found as well as
its behaviour when moving poles between the sheets. Our notations will be as follows:
in sections 3.1 we will describe the classical magnetic theory with electric couplings,
while in section 3.2 we will distinguish explicitly electric and magnetic
couplings.

\subsec{The Classical Vacua}

\vskip.02cm
\centerline{\it The Electric Theory}
\vskip0.2cm

To be definite we will consider the case in which only
one flavor, e.g. the last one, is higgsed. We will begin by considering the classical
theory with the simple KSS perturbation, namely the special case of \elth\ with
\eqn\simpel{
W_{el}={t_n\over n+1}\Tr X^{n+1}+m_2\tilde Q_{N_f}XQ^{N_f}+m_1\tilde Q_{N_f}Q^{N_f}.
}
This theory does not confine in the IR. Instead, the superpotential \simpel\
drives the flow to an interacting SCFT. The higgs vacuum in the electric theory is
obtained by giving a classical expectation value to the last flavor of fundamentals
\eqn\vevQ{
\tilde Q_{N_f\alpha}=(\tilde h,0,\ldots,0),\quad Q^{N_f}_\alpha=(h,0,\ldots,0),
}
then the quark equations of motion
fix the value of the adjoint to $X=\diag(x_1,0,\ldots,0)$ where $x_1=-m_1/m_2$.
The expectation value of the quarks
is fixed by the adjoint equations of motion to $h\tilde h=-t_nx_1^n/m_2$.

As usual, we can think of the low energy theory in two stages.
First, by higgsing the quarks we decrease the number of colors from $N_c$ to $N_c-1$. The quarks
$Q^\alpha_{N_f}$, $\tilde Q_\alpha^{N_f}$ for $\alpha=2,\ldots,N_c$ become the transverse component of
a massive vector superfield of mass squared $\tilde h h$,
while the components $X^1_\alpha$, $X_1^\alpha$ for $\alpha=2,\ldots,N_c$
of the adjoint replace the last flavor,
so the total number of flavors does not decrease.
Secondly, this latter new flavor acquires a mass $t_nx_1^{n-1}$ by expanding the adjoint superpotential.
The low energy theory is a $U(N_c-1)$ gauge theory with $N_f-1$ flavors.
The matching of the electric scales is
\eqn\elsca{
\Lambda^{2N_c-N_f}_{N_c,N_f}=
{(-x_1)\over m_2}\Lambda^{2(N_c-1)-(N_f-1)}_{N_c-1,N_f-1}.
}

\vskip.02cm
\centerline{\it The Magnetic Theory}
\vskip0.2cm

The magnetic theory corresponding to \simpel\ is defined by
\eqn\magne{
W_{mag}=-{t_n\over n+1}\Tr Y^{n+1}+
{t_n\over \mu^2}\sum_{j=1}^n P_j\tilde q Y^{n-j}q+
m_2(P_2)^{N_f}_{N_f}+m_1(P_1)^{N_f}_{N_f}.
}
Unlike the previous pseudoconfining case, this vacuum
is characterized by a nonvanishing classical expectation value
for the singlets, corresponding
to the electric higgsed quarks, that we classically match as
\eqn\pjhig{
(P_j)^{N_f}_{N_f}=\tilde Q_{N_f}X^{j-1}Q^{N_f}=-{t_nx_1^n\over m_2}x_1^{j-1},
}
for $j=1\ldots,n$. In this case the classical chiral ring is more complicated,
due to the nonvanishing singlets. In addition to the usual
singlet equations of motion \sina, we have also the quark
\eqn\quarks{
\sum_{j=1}^nP_j(Y^{n-j}q)=0,\quad\sum_{j=1}^nP_j(\tilde qY^{n-j})=0,
}
as well as the adjoint equations of motion.
%\eqn\yeom{
%Y^n={1\over \mu^2}\sum_{i=1}^{n-1}P_i\sum_{s=0}^{n-i-1}(\tilde q Y^s)(Y^{n-i-1-s}q).
%}
Nevertheless one can check that, in the convenient notation of \pseuY\ and \conadj,
the adjoint is in block diagonal form $Y=\diag(Y_{higgs},0,\ldots,0)$
and the nonvanishing part of the solution is
\eqn\maggii{\eqalign{
Y_{higgs}=&|1\rangle\langle v|+b_2 R_{n-1},\cr
|v\rangle=&-\sum_{j=1}^{n-1}x_1\left({x_1\over b_2}\right)^{j-1}|j\rangle,\cr
|\tilde q^{N_f}\rangle=&b_2|1\rangle,\qquad |q_{N_f}\rangle=b_2|n-1\rangle,
}}
where $b_2$ is given by \billi.
The first $(n-1)\times (n-1)$ block of the adjoint reads
\eqn\yhig{
Y_{higgs}=\pmatrix{
-{x_1}               &  b_2  &  0  &      &       &            \cr
-x_1({x_1\over b_2})  &  0  & b_2  &  .   &       &          \cr
     .                                &     &  0  &  .   &    .            \cr
     .                                &     &     &  .   &   b_2   &           \cr
-x_1({x_1\over b_2})^{n-2}               &  0  &  .  &  .   &   0
}}

Let us figure the low energy theory. First, by
higgsing the theory we break the gauge symmetry as $U(\oN_c)\to U(\oN_c-n+1)$
and  note that $\oN_c-n+1=n(N_f-1)-(N_c-1)$, as expected from the electric theory.
Accordingly, $\tilde q^{N_f}_\alpha$, $q_{N_f}^\alpha$, $Y^\alpha_m$, $Y_\alpha^s$ for $\alpha=n,\ldots,\oN_c$,
$m=2,\ldots, n-1$ and $s=1,\ldots,n-2$ conspire to join $n-1$ massive vector superfields in the fundamental
of $U(\oN_c-n+1)$ with mass $b_2$.
The flavor that disappears is replaced by a new flavor
$Y_1^\alpha$, $Y_\alpha^{n-1}$ for $\alpha=n,\ldots,\oN_c$,
so the number of flavors does not decrease.
Secondly, we look for a mass term for the new flavor coming from the superpotential
\eqn\masfla{
{t_n\over n+1}\Tr Y^{n+1}\simeq t_n(-x_1) b_2^{n-2}Y^\alpha_1 Y^1_\alpha,\quad \alpha=n,\ldots,\oN_c.
}
The number of flavors thus decreases by one unit also in the magnetic theory.
The matching of the magnetic scale goes as follows
\eqn\magmat{
\tilde{\Lambda}^{2\oN_c-N_f}_{\oN_c, N_f}={b_2^n\over t_n(-x_1) }
\tilde{\Lambda}^{2(\oN_c-n+1)-(N_f-1)}_{\oN_c-n+1,N_f-1}.
}
We can use the relation \scalerel\ between the scales
and find that this solution is consistent with the flows
\eqn\match{
\Lambda^{2(N_c-1)-(N_f-1)}_{N_c-1,N_f-1}
\tilde{\Lambda}^{2(\oN_c-n+1)-(N_f-1)}_{\oN_c-n+1,N_f-1}=\left({\mu^2\over t_n^2}\right)^{N_f-1}.
}

\vskip0.2cm
\centerline{\it Generic Polynomial Deformation}
\vskip0.2cm

We can generalize this to an arbitrary polynomial deformation
\eqn\simpip{
W_{el}=\Tr V(X)+m_2\tilde Q_{N_f}XQ^{N_f}+m_1\tilde Q_{N_f}Q^{N_f}.
}
The classical solution is the same
as in \vevQ\ the only difference being that now $\tilde h h =-V'(x_1)/ m_2$.

In the magnetic theory, the corresponding solution is as in \maggii\ but now the vector
$|v\rangle$ is replaced by
\eqn\generi{
|v'\rangle=-\sum_{j=1}^{n-1}\left({x_1\over b_2}\right)^{j-1}
\left(x_1+\sum_{l=0}^{j-1}x_1^{-l}{t_{n-l-1}\over t_n}\right)|j\rangle.
}

Now that we have the generic adjoint polynomial $V(z)$,
we can keep on flowing by further breaking the gauge group down to the low energy SQCD blocks.
The electric adjoint is then $X=\diag(x_1,a_1,\ldots,a_n)$, where $a_i$ are the roots of $V'(z)$
that appear with multiplicity $N_i$ such that $\sum_iN_i=N_c-1$.
In the magnetic theory we have correspondingly
a bunch of diagonal blocks $Y=\diag(Y_{higgs},Y_{a_1},\ldots,Y_{a_n})$,
where the first one is \generi\ and the others are as in \yai. In this vacuum
the relation between the low energy electric and magnetic gauge groups is
\eqn\lowni{
\oN_i=N_f-N_c-1,
}
since in the higgsed electric theory we have $\sum_{i=1}^n N_i=N_c-1$.
We can also compute the classical expression of the generators of the
chiral ring operators in this vacuum.
The resolvent $\tilde R(z)$ vanishes, while
\eqn\claschig{\eqalign{
\tilde M_{cl}(z)=&-\mu^2{m(z)\over V'(z)-V'(x_1)},\cr
\tilde T_{cl}(z)=&\sum_{i=1}^n{\oN_i\over z-a_1}+{d\over dz}\ln{V'(z)-V'(x_1)\over z-x_1}.
}}
Note that \claschig\ gives the correct behaviour at infinity $T_{cl}\sim \bar N_c/z$
since $\sum_{i=1}^n \bar N_i=\bar N_c-n+1$.

Let us mention that this description is agreement with the expectations from electric magnetic
duality. If we compare this solution to the pseudoconfining one \conadj,
we see that while the electric theory, being higgsed, becomes more weakly
coupled, in the magnetic theory the rank of the higgsed block in
the adjoint decreases from $n$ to $n-1$, thus making the theory more strongly coupled.

One can find a small generalization of the solution \maggii\
by turning on higher meson perturbations in the electric theory, always along
the last electric flavor direction.
In Appendix C we will give more details about the solutions with several higgsed electric
colors, but now let us add just few comments.
In this way we can have more higgsed entries in the same flavor
$Q^{N_f}=(h_1,\ldots,h_l,0,\ldots,0)$ and correspondingly
$X=\diag(x_1,\ldots,x_l,0,\ldots,0)$, the electric theory being at weaker coupling. The
general structure of the magnetic expectation values is that the first
$Y_{higgs}$ block decreases its rank down to $n-l$. Hence,
the magnetic side looks more strongly coupled. The rank of the generic Seiberg
blocks is still $\oN_i=N_f-N_i-1$ and one can check that still
\eqn\ranks{
\sum_{i=1}^n(N_f-N_i-1)+n-l=nN_f-N_c=\oN_c,
}
since now $\sum_{i=1}^n N_i=N_c-l$. We can carry on this procedure
until $l=n-1$: one further higgsing would get the rank of $Y_{higgs}$ vanished. In
fact $l=n-1$ is also the maximal number of Higgs eigenvalues we can turn
on on the same electric flavor, i.e. the largest value the degree of the meson
polynomial $m(z)$ can reach, higher mesons being trivial in the electric chiral ring.
Finally, if we allow for different electric flavors to get
higgsed then in the magnetic theory we have to add a new block analogous
to $Y_{higgs}$ for each higgsed flavor. We can not go on higgsing forever,
issues similar to the one that led our discussion of \nacchio\ arise
also in this phase.

\subsec{The Chiral Ring}

\vskip0.2cm
\centerline{\it The Electric Theory}
\vskip0.2cm

Let us consider the minimal case in which the electric theory admits a higgs vacuum
and we can safely apply the DV method: all flavors are massive and
a Yukawa interaction is turned on only for the last flavor. The tree
level superpotential is
\eqn\elhig{
W_{tree}=\Tr V(X)+m_1\tQ_{f} Q^{f}+m_2\tQ_{N_f}XQ^{N_f},
}
so that the meson polynomial reads $m(z)_{f}^{\tf}=m_1\delta_{f}^{\tf}+
z \,m_2\delta_f^{N_f}\delta^{\tf}_{N_f}$ and has only one root $x_1=-m_1/m_2$.
We give a classical expectation value to the last flavor of quarks and consider
the following solution to the equations of motion
\eqn\elvac{\eqalign{
X=&\diag(x_1,a_1,\ldots,a_n)\cr \tilde Q_{N_f}=&(\tilde
h_1,0,\ldots,0),\quad Q^{N_f}=(h_1,0,\ldots,0),
}}
where each $a_i$ is a root of $V'(z)$ and has multiplicity $N_i$ such that $\sum_{i=1}^n N_i=N_c-1$.
The adjoint equations of motion set $\tilde h_1 h_1 =-V'(x_1)/ m_2$.
In the classical chiral ring the resolvent $R(z)$ vanishes, while the nonvanishing
generators are
\eqn\claschiel{\eqalign{
T(z)|_{cl}=&{1\over z-x_1}+\sum_{i=1}^n{N_i\over z-a_i},\cr
M_{N_f}^{N_f}(z)|_{cl}=&-{V'(x_1)\over z-x_1}\oint_{x_1}{dx\over m_{N_f}^{N_f}(x)}=-{1\over m_2}{V'(x_1)\over z-x_1},
}}

Let us solve the anomaly equations. The resolvent $R(z)$ is always \solveR.
The story is different now for the generator of the mesons $M(z)$. The boundary conditions in the higgs
vacuum require a pole on the first sheet along the last flavor direction. The solution along the
pseudoconfining flavor directions is the usual one
\eqn\meselgea{
M(z)_f^{\tf}=R(z)m_1^{-1}\delta^{\tf}_f,\qquad (f,\tf)\neq (N_f,N_f),
}
while the solution along the last flavor direction is
\eqn\meselgehi{
M(z)_{N_f}^{N_f}={R(z)\over m_1+z\,m_2}-{V'(x_1)-R(x_1)\over z-x_1}m_2^{-1}.
}

We can integrate \meselgehi\ on the contour $A$ that encircles all the branch
points of the resolvent and obtain the quantum expressions for the mesons.
There are two types of mesons, the ones in the $(f,\tf)\neq(N_f,N_f)$
flavor directions that are exactly given by \elmesono, and the ones in the last flavor direction
that are
\eqn\lastme{
\tQ_{N_f}X^{j-1}Q^{N_f}=-\sum_{i=1}^n{a_i^{j-1}f(a_i)\over (m_1+a_im_2)V''(a_i)}+\ldots,
}
where the dots stand for higher terms in the semiclassical expansion of the resolvent \semifi.

\vskip0.2cm
\centerline{\it The Magnetic Theory and its Analytic Structure}
\vskip0.2cm

The magnetic theory corresponding to \elhig\ is defined by the following tree level
superpotential
\eqn\magtre{
W_{mag}=\Tr \oV(Y)+\tilde q^{\tilde f}\tilde m(Y)^f_{\tilde f} q_f+\om_1\tr P_1+\om_2 (P_2)_{N_f}^{N_f},
}
The anomaly equation for the resolvent $\tilde R(z)$ gives the usual solution \solver.
The equations for $\tilde M(z)$ and the singlet equations of motion
now have different boundary conditions depending on the flavor directions.
The first $(f,\tf)\neq (N_f,N_f)$ flavors have the
same solution \solPP\ and \solP\ as in the first massive case we considered,
in which $\tm(z)$ is proportional to the quantum deformation
\eqn\mimimmi{\eqalign{
\tilde M(z)^{\tf}_f=&\tilde R(z)\tilde m(z)^{-1}{}^{\tf}_f,\cr
\tilde m(z)_{\tf}^f=&-{\of(z)\over 4\mu^2}(\om^{-1})_{\tf}^f.
}}

The remaining flavor direction $(f,\tf)=(N_f,N_f)$ corresponds to the higgsed electric
meson. The new boundary conditions for $\tilde M(z)$ are
$n-1$
poles on the first sheet and no pole on the second sheet, as opposed
to the previous pseudoconfining case \genemsgen\ in which no pole
was there on the first sheet and $n-1$ poles appeared on the second sheet
\eqn\hiM{
\tilde M(z)^{N_f}_{N_f}={\tilde R(z)\over\tm(z)^{N_f}_{N_f}}
-\sum_{i=1}^{n-1}{\oV\,'(e_k)-\tilde R(e_k)\over z-e_k}{1\over\tm'(e_k)^{N_f}_{N_f}},
}
where $e_k$ for $k=1,\ldots,n-1$ are the roots of $\tilde m(z)$.

The picture of the analytic structure of $\tilde M(z)$ is the following. We saw that in the
pseudoconfining electric case \convac, the magnetic solution
\genemsgen\ does not have poles on the first sheet, but it has
$n-1$ poles on the second sheet. In the classical limit, these poles are very large, but in the quantum theory they move to the
region near the branch cuts, as we will check explicitly in
section 4.
In the electric higgs phase \vevQ, the magnetic solution \hiM\ gets $n-1$ poles on the first sheet
and no pole on the second sheet. The classical limit of this last solution has
still $n-1$ poles, coming from the second term in \hiM\ and the
fact that classically $\tm$ is nonvanishing.
Now let us move back to the electric theory and higgs two color direction on
the same electric flavor, replacing \elvac\ with
\eqn\twoco{
\tilde Q_{N_f}=(\tilde h_1,\tilde h_2,0,\ldots,0),\qquad
 Q^{N_f}=(h_1,h_2,0,\ldots,0).
}
and $X=\diag(x_1,x_2,0,\ldots,0)$.
The gauge group is higgsed down to $U(N_c-2)$ and the electric
theory becomes more weakly coupled. Classically we saw in \ranks\ that
the rank of the corresponding magnetic higgs block decreases by one. Quantum mechanically
this corresponds to moving one of the $n-1$ poles in \hiM\ from the first to the second sheet
\eqn\himi{
M(z)^{N_f}_{N_f}={R(z)\over\tm(z)}-\sum_{i=2}^{n-1}{V'(e_k)-R(e_k)\over z-e_k}{1\over\tm'(e_k)}
-{R(e_1)\over(z-e_1)}{1\over \tm'(e_1)}.
}
In this way the magnetic theory becomes more strongly coupled. In the classical
limit $\tilde m(z)$ is nonvanishing so we are left with just $n-2$ poles.
Note that we can
higgs at most $n-1$ electric color directions on the same flavor,
$Q^{N_f}=(h_1,\ldots,h_{n-1},0,\ldots,0)$, corresponding to the largest degree
the electric meson polynomial $m(z)$ can have. On the magnetic side, there are at most $n-1$ poles
to be moved all the way to the second sheet. When we pass them all, the corresponding meson generator
looks much like \genemsgen, but actually it is different. While the classical
limit of \genemsgen\ is nonzero due to the fact that $\tilde m(z)$ vanishes
classically, in this case $\tilde m(z)$ is always nonvanishing and
therefore the meson generator vanishes classically.

In the previous pseudoconfining case, we noted that, even if classically
$\tilde M(z)$ has some singularities, in the quantum theory
these singularities are smoothed out and we end up with a regular expression
in the first semiclassical sheet. In the higgs case, instead, the singularities
we might expect in the classical generator do not disappear at the quantum
level but are genuine poles in the quantum expressions \hiM.

We still have to fix $\tilde m(z)$ by requiring that the
singlet equations of motion \fineom\ are satisfied.
The contour $\tilde A$ in \fineom\ encircles all the branch points of the resolvent, but now
it encircles also the $n-1$ poles at $e_k$. The evaluation of this contour
integral is much more complicated than in the pseudoconfining
case \fineom, since we get additional
residues at $e_k$. Dropping higher terms in the semiclassical
expansion of the resolvent \semifi\ and showing just the leading approximation we get
\eqn\singlehi{
\sum_{i=1}^n\left[-{\oa_i^{l-1}\of(\oa_i)\over 4\tm(\oa_i)\oV\,''(\oa_i)}-
\mu^2{\oa_i^{l-1}\om(\oa_i)\over \oV\,''(\oa_i)}\right]+
\sum_{k=1}^{n-1}{2\tilde R(e_k)-\oV\,'(e_k)\over \tm'(e_k)}e_k^{l-1}=0,
}
for $l=1,\ldots,n$. Again we see that \singlehi\ amounts to $n$ conditions that
implicitely fix the unknown polynomial $\tm(z)$. However,
in this case it is hard to solve
these equations explicitly since the roots $e_k$ appear inside the resolvent.

Now consider the matching \matcha\ between the gauge singlets and the electric mesons.
The mesons in the directions $(f,\tf)\neq(N_f,N_f)$ match as in the first massive case,
reobtaining the map
\eqn\rema{
\oV\,'(z)=-V'(z),\qquad \of(z)=f(z),\qquad m^{\tf}_f=\om^{\tf}_f.
}
The last direction $(f,\tf)=(N_f,N_f)$ gives
a new condition, that we can write as
\eqn\macci{
\oint_{A'}z^{l-1}\left[M_{el}(z)^{N_f}_{N_f}+\mu^2{\tm(z)^{N_f}_{N_F}\over V'(z)}\right]=0,
}
for $l=1,\ldots,n$. The electric meson generator is given by \meselgehi\ and we
replaced the magnetic adjoint polynomial with the electric one by \rema.
The contour $A'$ now is a very large
contour that encircles the branch points of the resolvent
as well as the electric pole at the point $x_1$ in the first sheet.
Evaluating the contour integral at first order in the semiclassical expansion
and dropping the higher terms we find
\eqn\maccifi{
\sum_{i=1}^n{a_i^{l-1}f(a_i)\over 4m(a_i)V''(a_i)}-
{2R(x_1)-V'(x_1)\over m_2} x_1^{l-1}+\sum_{i=1}^n
\mu^2{a_i^{l-1}\tm(a_i)\over V''(a_i)}=0,
}
for $l=1,\ldots,n$.

Had we not allowed the contour to encircle the pole at
$x_1$, this expression would have had an inconsistent classical limit. Let us consider in fact the
classical limit of the conditions we have found so far. This is achieved by
setting to zero the quantum deformation $f(z)$ so that the resolvent vanishes in the first sheet.
It is more transparent to write the two classical conditions as contour integrals.
We fix the classical polynomial $\tm_{cl}(z)$
by the singlet equations
\eqn\thea{
\oint_{A_{cl}} z^{l-1}\left[\mu^2{m(z)\over V'(z)}+{V'(z)\over \tm(z)}\right]=0,
}
for $l=1,\ldots,n$, where the contour encircles all the poles of the two meromorphic functions.
By picking up the residues we get
\eqn\classi{
\mu^2\sum_{i=1}^n{a_i^{l-1}m(a_i)\over V''(a_i)}+\sum_{k=1}^n\hat{e}_k^{l-1}{V'(\hat{e}_k)\over
\tm'_{cl}(\hat{e}_k)}=0,
}
where we hatted the classical roots $\hat e_k$. This condition is much easier
to solve than \singlehi\ due to the disappearance of the resolvent.
Once we fix $\tm_{cl}(z)$, the classical limit of the matching condition \macci\ is satisfied
\eqn\theb{
\oint_{A'_{cl}} z^{l-1}\left[\mu^2{\tm(z)\over V'(z)}+{V'(z)\over m(z)}\right]=0,
}
for $l=1,\ldots,n$, whose evaluation yields
\eqn\clamacci{
\mu^2\sum_{i=1}^n{a_i^{l-1}\tm_{cl}(a_i)\over V''(a_i)}+x_1^{l-1}V'(x_1)m_2^{-1}=0.
}

\newsec{The Cubic Superpotential}

In this section we will illustrate the pseudoconfining and higgs phase computations,
worked out in the previous sections, in the simplest example
that allows for a higgs phase, namely a cubic interaction for the adjoint.

\subsec{The Pseudoconfining Case}

Let us consider an electric tree level superpotential as in \elhig\ and let us specialize
to $n=2$. We take the following adjoint polynomial
$$
V'(z)=t_1z+t_2z^2,
$$
whose roots are $a_1=0$ and $a_2=-t_1/t_2$. We also have a meson polynomial $m(z)^{\tf}_f
=m_1\delta^{\tf}_f+z\,m_2\delta^{N_f}_f\delta^{\tf}_{N_f}$. The resolvent is
$2R(z)=V'(z)-\sqrt{V'(z)^2+f(z)}$ and its quantum deformation is
$f(z)=f_0+f_1z$.

In the magnetic theory, all the flavor directions $(f,\tf)\neq(N_f,N_f)$ correspond
to the massive case solved in section 2.4. In the following we will focus instead
on the last direction $(f,\tf)=(N_f,N_f)$ only and suppress the flavor indices.
We will see an explicit example of
the computations in section 2.7. Let us consider
$\tilde m(z)=\tilde m_1+\tilde m_2 z$, whose one root we denote as
$e_1=-\tm_1/\tm_2$. The solution \conje\ of the magnetic theory is given
by the condition $4\mu^2 m(a_i)\tm(a_i)=f(a_i)$ for $i=1,2$, from which we
get the $\tilde m(z)$ coefficients in terms of $m(z)$ and $f(z)$
\eqn\coefti{\eqalign{
\tilde m_1=&{f_0\over4\mu^2m_1},\cr
\tilde m_2=&-{t_2\over 4\mu^2t_1}\left({t_2f_0-t_1f_1\over t_2m_1-t_1m_2}-{f_0\over m_1}\right),
}}
so that the singlet equations of motion \sina\ are satisfied. This condition also ensures
that the singlets $P_j$ extracted from \singlets\ match the electric mesons
$\tQ X^{j-1}Q$.

We would like to check
that the root $e_1$ lies inside the contour $\tilde A$ that encircles the branch points of the resolvent.
Consider the classical limit of this setup. In this
limit both $f(z)$ and $\tilde m(z)$ vanish, but we still have to satisfy the singlet equations of motion.
We first want to obtain the dependence of $f(z)$ on the total glueball and then perform the limit
by sending the glueball to zero. For this purpose we have to choose
a vacuum for the electric theory and solve the factorization of gauge theory curve.
Let us consider the phase
in which the gauge group is unbroken, which corresponds to the one--cut case, namely
the electric adjoint is $X=\diag(a_i,\ldots,a_i)$. Then the
curve factorizes as
\eqn\facto{
V'(z)^2+f(z)=t_2^2(z-k)^2(z-a+b)(z-a-b),
}
with one double root and two branch points. We already know from \solveS\
that $f_1=-4t_2S$ and we can find \aldacira\
\eqn\solfa{\eqalign{
k=&-{t_1\over t_2}+a,\cr
a=&{t_2\over t_1} S +{\cal O}(S^2),\cr
b=&\sqrt{S\over 2m}\left(2+{\cal O}(S)\right).
}}
We don't need the full result, but just the leading terms in the glueball, from
which we find $f_0=-2t_1S+{\cal O}(S^2)$.
Then in the classical limit $S\to0$ we have
\eqn\fzerone{
{f_0\over f_1}\sim{t_1\over t_2}+{\cal O}(S).
}
The root $e_1$ of $\tilde m(z)$ in the classical limit is
\eqn\rootil{
\hat{e}_1={t_1\over t_2}{t_2m_1-t_1m_2\over t_1m_2-2t_2m_1}.
}
In the limit of large mass $m_1$ for the electric quarks,
we find
$\hat{e}_1\sim-{t_1\over 2t_2}$, which is not large but
lie inside the contour $\tilde A$ that encircles the branch points of the
resolvent, as we claimed below Eq.\fineom.
In particular, this classical pole is halfway between the two roots of $V'(z)$.

\subsec{The Higgs Case}

We keep the same superpotential, but consider now the
electric higgs vacuum $X=\diag(x_1,0,\ldots,0)$ and
\eqn\elicu{
\tilde Q_{N_f}=(\tilde h_1,0,\ldots,0),\qquad
Q^{N_f}=(h_1,0,\ldots,0),
}
where the gauge group is higgsed down to $U(N_c-1)$ and the electric equations of motion
set $\tilde h_1h_1=-V'(x_1)/m_2$.
This vacuum is characterized by a nonvanishing classical expectation value
for the electric mesons
\eqn\clamecu{
\tilde Q_{N_f}X^{j-1}Q^{N_f}=(P_j)^{N_f}_{N_f}=-{x_1^{j-1}V'(x_1)\over m_2}.
}

We want to check the prescription we outlined in section 3.2 in the magnetic theory.
In the higgs phase the singlets $P(z)$ as well as the magnetic polynomial
$\tm(z)$ acquire a classical expectation value. From \clamecu\ we can read out
their classical expressions
\eqn\clapimi{\eqalign{
P(z)_{cl}=&{\oV\,'(\bar x_1)\over \om_2z^2}(z+\bar x_1),\cr
\tm(z)_{cl}=&{\oV\,'(\bar x_1)\over \mu^2 \bar m_2}(t_1+t_2\bar x_1+zt_2).
}}

Now we would like to solve the quantum theory at first order
in the semiclassical expansion. If we look at the flavor directions
$(f,\tf)\neq(N_f,N_f)$ we find the duality map \rema, that we can use
in the following computation.
In the higgsed direction, first we have to solve the singlet equations of motion \singlehi\
and then check that the matching relation \maccifi\ is satisfied. But this is kind of hard,
due to the presence of the resolvent in the last term of \singlehi\ that makes the equations
pretty much involved. However, since the solutions of \maccifi\ must be solutions of \singlehi\ too,
the best we can do is we solve the matching condition \maccifi\ and then try to check that this solution
satisfies the singlet equations of motion \singlehi, thus getting it the other
way around.

The matching condition at first order is
\eqn\maccicu{
\sum_{i=1}^2\left[-{a_i^{l-1}f(a_i)\over 4\om(a_i)V''(a_i)}-
\mu^2{a_i^{l-1}\tm(a_i)\over V''(a_i)}\right]+
{2R(x_1)-V'(x_1)\over \om_2} x_1^{l-1}=0,
}
for $l=1,2$. This can be solved easily with the result
\eqn\solmacu{\eqalign{
\tilde m_1=&-{f_0-4V'(\bar x_1)[V'(\bar x_1)-2R(\bar x_1)]\over 4\mu^2\bar m_1},\cr
\tm_2=&{t_2\over 4\mu^2\bar m_2 V'(\bar x_1)}
\left[f(\bar x_1)-4V'(\bar x_1)[V'(\bar x_1)-2R(\bar x_1)]\right].
}}
Quantum mechanically, the one root of $\tm(z)$ is $e_1=-\tm_1/\tm_2$. If we take the classical
limit of \solmacu\ we obtain the expected expression \clapimi\ and
its classical root $\hat{e}_1=-x_1-t_1/t_2$, by identifying $\bar m(z)=m(z)$. In the semiclassical electric
picture in which the higgs vev $x_1$ is large, this root gets very large, too.
This phase is very different from \rootil, where for large electric quark masses
we got small $\hat{e}_1$.

It would be very hard to check that \solmacu\ satisfies the singlet equations of motion \singlehi\
at first order in the semiclassical expansion, due to the fact that the
resolvent should be evaluated at the root of $\tm(z)$. But one can still easily check that indeed the classical
limit of the singlet equations
\eqn\classicu{
\mu^2\sum_{i=1}^2{a_i^{l-1}m(a_i)\over V''(a_i)}+\hat{e}_1^{l-1}{V'(\hat{e}_1)\over
\tm'_{cl}(\hat{e}_1)}=0,\qquad l=1,2,
}
is satisfied by $\hat{e}_1$ and the classical limit of \solmacu.

\newsec{Discussion}

Let us summarize our results
and suggest some further speculations. At the classical level, we generalized the KSS
solution to the case of polynomial superpotentials, allowing
for generic meson deformations, and we found the
solutions of the magnetic theory corresponding to the electric
pseudoconfining and higgs vacua.
We considered then duality in the quantum theory and we used the DV approach to
solve for the chiral rings just above the mass gap: we studied the effective glueball
superpotential. We analyzed the following
three cases:
\item{1.} The electric meson superpotential is a mass term for all the flavors.
We saw that electric--magnetic duality holds exactly offshell in this case.
\item{2.} The generic pseudoconfining phase, where we allow for a generic
meson deformation, has a way richer analytic structure.
We matched the electric mesons with the magnetic singlets at first order
in the semiclassical expansion of the resolvent. In this way we found
a condition that reproduces the Konishi anomaly equation in the low energy SQCD blocks
and their magnetic dual. In this case duality does not hold exactly offshell.
\item{3.} In the electric higgs phase, we found the solution to
the magnetic theory, at first order in the semiclassical
expansion, and showed that it is consistent with the classical limit.
Neither in this case does duality work exactly offshell.
Moreover, while in the pseudoconfining case the classical singularities in
$\tilde M(z)$ are just an artifact of the classical solution
and in the quantum theory they disappear, in the higgs case the classical singularities are preserved in the
quantum theory.

As we just summarized,
electric--magnetic duality holds exactly offshell in some cases,
namely in SQCD or when the meson
polynomial $m(z)$ is $z$--independent,
while in more general cases it should work exactly only onshell.
It would be nice to find a physical motivation for such different behaviors.

We could draw a picture of the
analytic properties of the magnetic theory as we continuously interpolate between different
higgs vacua in the electric theory (when we move poles
from the second to the first electric sheet).
An interesting extension of our analysis would be to show what happens on the magnetic
side when we smoothly pass from the pseudoconfining to the higgs phase in
the electric theory. In this way, one might shed some light on the onshell process
that takes place when a branch cut of the resolvent closes up,
as recently investigated in \masaki. On the electric side this
is a strong coupling phenomenon, but one should describe it easily
in the dual regime.

On the other hand, it would be interesting to use our quantum
duality map to gain insight on the meaning of the
electric parameter $L$ introduced in \CSW\ as the degree of the determinant
of the meson polynomial $B(z)=\det m(z)$. This parameter plays the role
of an effective number of flavors and is related to the appearence of
instanton corrections to the classical chiral ring. In particular,
if the electric superpotential $V(z)$ has degree $N_c+1$, when $L\geq N_c$ the strong coupling analysis
shows that the classical
Casimirs $\Tr X^j$ for $j=1,\ldots,N_c$ are modified in the quantum chiral ring
by terms proportional to the instanton factor. It would
be interesting to understand the corresponding phenomenon in the magnetic theory.
In our setup, $L\leq N_f(n-1)$, so the condition for the appearence
of instanton corrections is related to $N_f\leq\oN_c$ on the magnetic side.

A natural generalization of our analysis would be to consider
$SO(N_c)$ and $Sp(2N_c)$ gauge groups. In particular, one could
translate into a magnetic language the
map between $Sp(2N_c)$ theory with an antisymmetric tensor and $U(2N_c+2n)$ with an adjoint,
recently proposed in \cachazo. Moreover, the KSS duality
has been generalized to theories with two adjoint chiral superfields and
fundamentals in \JB. It would be nice to extend our classical and
quantum mechanical solutions to this case. One might find
some unusual features, due to the fact that the gauge theory curve
is not hyperelliptic anymore in this theory and
cannot be obtained as a deformation of an $\cN=2$ theory.

We would like to make one last remark on the theory without superpotential,
whose magnetic dual is not known.
In \KSS\ it was suggested that one might try to obtain this theory as
a certain limit of the KSS theory with superpotential $t_n \Tr X^{n+1}$.
Since the limit of vanishing $t_n$ is singular,
it was suggested to study
the $k\to\infty$ limit instead, so that the magnetic dual might look
like an $U(\infty)$ gauge theory, which is expected to behave like a string theory.
The story might be simpler, though. Due to the recent work of Intriligator and Wecht
\IW, we know
that an analogue of the conformal window of SQCD exists also for the KSS theory:
it is the region in the range of $N_f$ in which both
the electric and the magnetic deformations $\Tr X^{n+1}$ and $\Tr Y^{n+1}$ are
relevant \KPS. Now,
if we take a sufficiently large number of flavors we can make
the deformation $\Tr X^{n+1}$ irrelevant, but still keeping
the electric theory asymptotically free.
Therefore, the electric theory at the fixed point will be
the theory without superpotential. But on the magnetic side, the corresponding
superpotential keeps being relevant and we have the usual full magnetic theory.
So we might not really need to take $k$ very large to
remove the electric superpotential, hence the magnetic dual
of the theory without superpotential need not be a kind of string theory.
This point might deserve further study.

\vskip0.2cm
\centerline{\bf Acknowledgements}

I would like to thank A. Schwimmer for help at
every stage of this project.
It is also a pleasure to thank L. Alday, M. Bertolini, L. Bonora, F. Cachazo,
M. Cirafici, J. David, M. Matone
and especially K. Intriligator and N. Seiberg for
discussions and suggestions. I would like to thank the organizers of the PiTP 2004 at the Institute for Advanced Studies, Princeton, for their kind hospitality and for providing a very stimulating environment, where this work has been completed.

\appendix{A}{The DV Method}

In the following we will quote some results on the generalized Konishi
anomaly approach to DV, following \CDSW\nati\CSW. We will collect some useful
formulae we need in the main part of the paper. For a review of the
DV approach and an extensive discussion of the huge literature
available by now see \Ferretti\ and references therein.

Consider the electric theory with tree level superpotential \elth\ and
specialize to a diagonal meson polynomial $m(z)^{\tf}_f=m(z)\delta^{\tf}_f$
of degree $l\leq n-1$. Classically,
we can consider the following generic vacuum
\eqn\vacchi{\eqalign{
X=&\diag(x_1,\ldots,x_{i},a_1,\ldots,a_n),\cr
\tQ_{N_f}=&(\tilde h_1,\ldots,\tilde h_{i},0,\ldots,0),\cr
\qquad Q^{N_f}=&(h_1,\ldots,h_{i},0,\ldots,0)
}}
where $i\leq l$ and the adjoint equations of motion fix $\tilde h_i h_i=-V'(x_i)/m'(x_i)$.
We will higgs only the last flavor of quarks for at most $l$ color directions.
We will introduce the occupation number $r_I$ for each root $x_k$ of the meson
polynomial: $r_k=1$ if the root appears in the adjoint in \vacchi\ and $r_k=0$ otherwise.
We will be interested in the chiral operators \chiralopf. Their classical
expressions are
\eqn\puppo{\eqalign{
M(z)_{cl}=&-\sum_{k=1}^{l}{r_kV'(x_k)\over z-x_k}{1\over m'(x_k)},\cr
T(z)_{cl}=&\sum_{i=1}^n{N_i\over z-a_i}+\sum_{k=1}^l{r_k\over z-x_k},
}}
while the resolvent $R(z)$ vanishes. In the first equation in \puppo, the occupation
number $r_k$ always vanishes unless $(f,\tf)=(N_f,N_f)$.

By considering a generalized version of the Konishi anomaly and the factorization
property of the chiral ring, we can write down the following algebraic equations for the generators
of the chiral ring \chiralopf\
\eqn\creosoto{\eqalign{
[V'(z)R(z)]_-=&R(z)^2,\cr
[M(z)m(z)]_-=&R(z),\cr
[V'(z)T(z)]_-+\tr[m'(z)M(z)]_-=&2R(z)T(z),
}}
where the subscript means that we are dropping the non--negative powers in the Laurent
expansion. By solving the first equation we learn that
\eqn\puppia{
2R(z)=V'(z)-\sqrt{V'(z)^2+f(z)},
}
where $f(z)$ is a degree $n-1$ polynomial with vanishing classical limit. This defines the curve
of the gauge theory to be the hyperelliptic Riemann surface $y^2=V'(z)^2+f(z)$, which
is a double--sheeted cover of the plane.

The solution of the second equation in \creosoto\ gives the meson generator $M(z)$, whose boundary conditions
depend on the classical vacuum \vacchi\ we have chosen
\eqn\panulna{
M(z)={R(z)\over m(z)}-\sum_{k=1}^l{r_kV'(x_k)+(1-2r_k)R(x_k)\over z-x_k}{1\over m'(x_k)},
}
where we suppressed the flavor indices, but we keep in mind
that $r_k$ can be nonzero only along the $(f,\tf)=(N_f,N_f)$ flavor direction.
The picture that emerges from \panulna\ is that an higgs eigenvalue in \vacchi, i.e. $r_k=1$, corresponds
in the quantum theory to a pole for $M(z)$ in the first semiclassical sheet of the curve at $x_k$,
while whenever $r_k=0$ we have a pole in the second sheet.

The solution to the third equation in \creosoto\ again depends on the boundary conditions \vacchi
\eqn\bucchia{
T(z)=\sum_{k=1}^l{1\over 2(z-x_k)}-\sum_{k=1}^l{(1-2r_k)y(x_k)\over2y(z)(z-x_k)}+{c(z)\over y(z)},
}
where $c(z)$ is another degree $n-1$ polynomial.

\appendix{B}{Some Properties of the Effective Glueball Superpotential}

In this Appendix we will consider some properties of the coefficients $h_i$
introduced in \bicon\ and, by using these expressions, we will prove \as\ up
to an assumption of integrability.

\subsec{Properties of the $h_i$}

Consider the
function $V(a_i)$ of the couplings $t_j$ defined as
\eqn\wtj{
V(a_i)=\sum_{j=1}^n{t_j\over j+1}a_i^{j+1},
}
where $a_i$ is solution of $V'(a_i)=0$. Note that we are considering the $a_i=a_i(t_j)$
as functions of the couplings.
Taking a derivative of $V(a_i)$ with respect to $t_k$ we obtain
\eqn\der{
{\partial V(a_i) \over \partial t_k}={a_i^{k+1} \over k+1},
}
the second term in taking the derivative vanishing since it is multiplied
by $V'(a_i)$.
Since \der\ is a derivative, it fulfills the condition
\eqn\pKN{
{\partial \over \partial t_l}{a_i^{k+1} \over k+1}=
{\partial \over \partial t_k}{a_i^{l+1} \over l+1},
}
and therefore
\eqn\KN{
{\partial \over \partial t_{n-l}} {a_i^{j+l} \over j+l} =
{\partial \over \partial t_{n-k}} {a_i^{j+k} \over j+k}.
}
which is our classical integrability condition.

Now we will assume that also the effective superpotential \as\
satisfies the integrability
condition
\eqn\interposto{
{\partial^2{\cal W}_{eff}\over \partial t_l\partial t_j}=
{\partial^2{\cal W}_{eff}\over \partial t_j\partial t_l}.
}
By using the classical integrability \KN, we find the relation
\eqn\intebi{
\sum_{i=1}^n{\partial h_i\over \partial t_l}{a_i^{j+1}\over j+1}=
\sum_{i=1}^n{\partial h_i\over \partial t_j}{a_i^{l+1}\over l+1}.
}
Note that this relation will hold also for the $d_i$ defined in \dideffi
\eqn\intedi{
\sum_{i=1}^n{\partial d_i\over \partial t_l}{a_i^{j+1}\over j+1}=
\sum_{i=1}^n{\partial d_i\over \partial t_j}{a_i^{l+1}\over l+1}.
}

Finally, let us consider a scaling argument on the coefficients $h_i=h_i(t_k,N_l,S_j)$.
Since
\eqn\Ndef{
N_c=\sum_{i=1}^nh_i=\oint_A{c(z)\over \sqrt{V'(z)^2+f(z)}},
}
if we rescale the glueballs $S_i\to\lambda S_i$ and the couplings $t_k\to\lambda t_k$,
we have correspondingly that $V'(z)\to \lambda V'(z)$ and $f(z)\to\lambda^2f(z)$,
while the $N_i$ are unchanged in \contin. But by \Ndef\ also the $h_i$ are invariant under the
scaling, meaning that they are homogeneous functions of the couplings and the glueballs
\eqn\homobi{
\sum_{i=1}^n\left(t_i{\partial \over \partial t_i}+S_i{\partial\over\partial S_i}\right)h_l=0,
}
and this property carries on to the $d_i$.

\subsec{Evaluation of ${\cal W}_{eff}$}

We will prove, up to the assumption \interposto, that, in the notations of section 2.5, if we define
\eqn\apptildab{
{\cal W}_{eff}=\sum_{i=1}^n d_iV(a_i),
}
then we have
\eqn\prooffi{
{\partial{\cal W}_{eff}\over \partial t_j}={1\over j+1}\sum_{i=1}^n h_i a_i^{j+1}.
}
Let us differentiate \apptildab\
\eqn\onepr{
{\partial {\cal W}_{eff}\over \partial t_j}=\sum_{i=1}^n{d_i\over j+1}a_i^{j+1}
+\sum_{i,k=1}^nt_k{\partial d_i\over \partial t_j}{a_i^{k+1}\over k+1}+
\sum_{i=1}^nd_i{\partial a_i\over \partial t_j}\sum_{k=1}^nt_ka_i^k,
}
but the last term vanishes since $V'(a_i)=0$. Now we need to evaluate $\partial d_i/\partial t_j$.
First note that $\partial_y d_i=d_i-h_i$ where $y=\sum_{i=1}^n\log S_i$. Then
by using the homogeneity \homobi\ of the $d_i$ we have that
\eqn\detidi{
\sum_{k=1}^n t_k{\partial d_i\over\partial t_k}=-\partial_y d_i.
}
Then we can use the integrability condition \intedi\ for the second term in \onepr\ and get
\eqn\finalprot{\eqalign{
{\partial{\cal W}_{eff}\over \partial t_j}=&\sum_{i=1}^n d_i{a_i^{j+1}\over j+1}-
\sum_{i=1}^n {\partial d_i\over\partial y}{a_i^{j+1}\over j+1}\cr
=&\sum_{i=1}^n d_i{a_i^{j+1}\over j+1}-\sum_{i=1}^n (d_i-h_i){a_i^{j+1}\over j+1}\cr
=&\sum_{i=1}^n h_i{a_i^{j+1}\over j+1}.
}}

\appendix{C}{Several Higgs Solution}

In this Appendix we will generalize the higgs solution \maggii\ to
the case in which more than one electric color direction is higgsed on the
same electric flavor.

\subsec{Two--Higgs case}

\vskip0.2cm
\centerline{\it The Electric Theory}
\vskip0.2cm

Consider the electric theory with superpotential
\eqn\birba{ W_{el}={t_n\over n+1}\Tr
X^{n+1}+ m_3\tilde Q_{N_f} X^2Q^{N_f} +m_2\tilde Q_{N_f}
XQ^{N_f}+m_1 Q_{N_f}  Q^{N_f}. }
We can get a classical vacuum in which the gauge group is higgsed as
$U(N_c)\to U(N_c-2)$ by considering the following expectation values
\eqn\soltwo{\eqalign{
X=&\diag(x_1,x_2,0,\ldots,0)\cr \tilde Q_{N_f}=&(\tilde
h_1,\tilde h_2,0,\ldots,0),\quad Q^{N_f}=(h_1,h_2,0,\ldots,0), }
}
where $\tilde h_i h_i =-V'(x_i)/ m'(x_i)$.
We denoted by $x_{1,2}$ the two roots of the meson polynomial
$m_3z^2+m_2z+m_1$. As is well known, the roots of a quadratic algebraic
equation satisfy
\eqn\cruu{ -(x_1+x_2)={m_2\over m_3},\qquad
x_1x_2={m_1\over m_3}. }

\vskip0.2cm
\centerline{\it The Magnetic Theory}
\vskip0.2cm

The superpotential for the magnetic theory is
\eqn\birbatti{
W_{mag}=-{t_n\over n+1}\Tr
X^{n+1}+ \tq\tm(P,Y)q+m_1(P_1)^{N_f}_{N_f} +m_2(P_2)_{N_f}^{N_f}
+m_3(P_3)_{N_f}^{N_f}.
}
By \match, the singlets acquire the classical expectation value
$P_j=\tilde h_1 h_1 x_1^{j-1}+\tilde h_2 h_2 x_2^{j-1}$ corresponding
to the electric mesons.

We expected the magnetic gauge group to break down to
$U(\bar N_c)\to U(\bar N_c-n+2)$, so that the vev for the adjoint
will be a nonvanishing block of rank $n-2$.
By using the property \cruu\ of the roots of $m(x)$ we can find
the solution to the singlet, fundamental and adjoint
equations of motion. In the notations of \conadj, the only nonvanishing entries in the adjoint are
\eqn\adjotwo{\eqalign{
Y=&|1\rangle\langle v_2|+b_3 R_{n-2},\cr
| v_2\rangle=&-\sum_{j=1}^{n-2}{1\over b_3}^{j-1}\Bigl[\left(-{m_2\over m_3}\right)^j
+\sum_{k=1}^{[j/2]}(j-k)\left(-{m_2\over m_3}\right)^{j-2k}\left(-{m_1\over m_3}\right)^k\cr
&+\delta^j_{even}\left({j\over2}\right)\left(-{m_1\over m_3}\right)^{[j/2]}\Bigr]|j\rangle,
}}
while the fundamentals are $|q_{N_f}\rangle=b_3|n-2\rangle$ and
$|\tilde q^{N_f}\rangle=b_3|1\rangle$ and $b_3$ is defined in \billi.

\subsec{Several Higgs}

\vskip0.2cm
\centerline{\it The Electric Theory}
\vskip0.2cm

Consider the electric theory with a generic meson polynomial on the last flavor
\eqn\birba{\eqalign{
W_{el}=&{t_n\over n+1}\Tr
X^{n+1}+ \tilde Q_{N_f} m(X)Q^{N_f},\cr
m(x)=&\sum_{k=1}^{l+1}m_k x^{k-1}. }}
The polynomial $m(x)$ has $l$ roots that we denote $x_1,\ldots,x_l$. The following property between the
coefficients of the polynomial and its roots holds
\eqn\grossadritta{
{m_{l+1-i}\over m_{l+1}}=(-)^i\sum_{k_1<k_2<\ldots<k_i}x_{k_1}x_{k_2}\ldots x_{k_i}.
}
We can consider the following vacuum
\eqn\solman{\eqalign{
X=&\diag(x_1,x_2,\ldots,x_l,0,\ldots,0)\cr
\tilde Q_{N_f}=&(\tilde
h_1,\tilde h_2,\ldots,\tilde h_l,0,\ldots,0),\quad Q^{N_f}=(h_1,h_2,\ldots,h_l,0,\ldots,0),
}}
where $\tilde h_i h_i =-V'(x_i)/m'(x_i)$. Note that each root $x_k$
can appear just once in the adjoint expectation value.
In this way we break the gauge symmetry as $U(N_c)\to U(N_c-l)$.
We can higgs at most $n-1$ colors on the same flavor, corresponding to the largest degree
the meson polynomial $m(z)$ can have.

\vskip0.2cm
\centerline{\it The Magnetic Theory}
\vskip0.2cm

According to the above discussion, in the magnetic theory we will have to solve
the singlet equations of motion
\eqn\sineom{
\eqalign{
\tilde q^{N_f}Y^{n-i}q_{N_f}=&{m_i\over m_{l+1}}b_{l+1}^{n-l+1},\qquad i=1,\ldots,l+1,\cr
\tilde q^{N_f}Y^{j}q_{N_f}=&0,\quad j=0,\ldots,n-l-2,
}}
where we used \grossadritta\ to ease the notation and $b_{l+1}$ is defined in \billi.
The solution for the adjoint, which generalizes \adjotwo,
can be sketched as the nonvanishing block of rank $n-l$
\eqn\manyhi{\eqalign{
Y=&|1\rangle\langle v_l|+b_{l+1} R_{n-l},\cr
|v_l\rangle=&-\sum_{j=1}^{n-l}{1\over b_{l+1}}^{j-1}\left[\left(-{m_l\over m_{l+1}}\right)^j+\ldots
\right]|j\rangle,\cr
|q_{N_f}\rangle=&b_{l+1}|n-l\rangle,\qquad |\tilde q^{N_f}\rangle=b_{l+1}|1\rangle,
}}
where the dots stand for an expression analogous to the one in \adjotwo\
but more involved.
In this way we break the magnetic gauge group down to
$U(\bar N_c -n+l)=U(n(N_f-1)-(N_c-l))$.
Note that this solution holds only for $l\leq n-1$, as we saw
on the electric side.

\listrefs

\end